\documentclass[twocolumn,
prb,preprintnumbers,amsmath,amssymb,floatfix]{revtex4}

\usepackage[linktocpage,bookmarksopen,bookmarksnumbered]{hyperref}
\usepackage{graphicx}
\usepackage{dcolumn}

\usepackage{amsmath,graphics,epsfig,color,verbatim,ulem}

\begin{document}

\setlength{\pdfpageheight}{\paperheight}
\setlength{\pdfpagewidth}{\paperwidth}

\title{Correlation-enhanced electron-phonon coupling: Applications of GW and \textit{screened} hybrid functional to 
bismuthates, chloronitrides, and other high T$_C$ superconductors}
\author{Z. P. Yin}
\email{yinzping@physics.rutgers.edu}
\author{A. Kutepov}
\author{G. Kotliar}
\affiliation{Department of Physics and Astronomy, Rutgers University, Piscataway, New Jersey 08854, United States.}
\date{\today}

\begin{abstract}
We show that the electron-phonon coupling (EPC) in many materials 
can be significantly underestimated by the standard density functional theory (DFT) in the 
local density approximation (LDA) due to large non-local correlation effects.
We present a simple yet efficient methodology to evaluate the realistic EPC going beyond LDA 
by using more advanced and accurate GW and \textit{screened} hybrid functional DFT approaches.
The corrections we propose explain the extraordinarily high superconducting temperatures 
that are observed in two distinct classes of compounds-the bismuthates and 
the transition metal chloronitrides, thus solving a thirty-year-old puzzle.
Our work calls for the critically reevaluation of the EPC of certain phonon modes in 
many other materials such as cuprates and iron-based superconductors.  
The proposed methodology can be used to design new correlation-enhanced high temperature superconductors 
and other functional materials involving electron-phonon interaction.
\end{abstract}
\maketitle

\section{Introduction}

The microscopic origin of superconductivity
has been established in numerous classes of materials. 
In conventional superconductors such as 
elemental metals it results from the exchange of phonons. In the
copper oxides, the iron pnictides and some organic compounds,
superconductivity is connected to close proximity to a magnetic state.
On the other hand, the cause of superconductivity in a large third
class of high temperature superconductors including
Ba$_{1-x}$K$_x$BiO$_3$ ( $T_{c,max}\simeq$32 K)\cite{Cava}, electron-doped
$\beta$-HfNCl compounds ($T_{c,max}\simeq$25.5 K)\cite{Yamanaka}, and
alkali-doped fullerides $A_3$C$_{60}$ (T$_{c,max}$ up to 40
K)\cite{Gunnarsson} is still mysterious. They have been referred
to as the ``other high-temperature superconductors",\cite{pickett2, Pickett} 
because neither spin nor phonon seems to be responsible for 
their high superconducting temperatures.

Since they are apparently diamagnetic (thus the spin degree of freedom seems irrelevant),
two charge mediated pairing mechanisms have been proposed (and ruled out) 
for the bismuthates: the ``negative U" scenario \cite{Taraphder,
varma} and strong coupling to an O-breathing phonon
mode\cite{wjin, rice}. The first scenario involves the tendency
of Bi atoms to avoid the Bi$^{4+}$ state, while preferring
the Bi$^{3+}$ and Bi$^{5+}$ charge state, thus generating an attractive local
``negative U" pairing center. However, first
principles calculations have so far ruled out the negative U
mechanism\cite{Vielsack, Harrison}.
Regarding the electron-phonon mechanism, 
previous approaches to the \textit{ab initio} evaluation of the
electron-phonon coupling (EPC) were based on the density
functional theory (DFT) and linear response theory (LRT)\cite{lmtart1}
in the local density approximation (LDA) or generalized gradient
approximation (GGA). 
The EPC calculated by this standard DFT-LDA/GGA approach 
is insufficient to account for the high T$_c$ in both 
Ba$_{1-x}$K$_x$BiO$_3$ ~\cite{Meregalli}  and the 
electron-doped transition metal chloronitrides\cite{ZrNCl, SCDFT} 
within the Migdal-Eliashberg theory,
which seems to rule out the phonon mechanism.

However, to rule out the phonon mechanism, 
the EPC has to be evaluated reliably.
It is well-known that the standard DFT-LDA approach fails to 
describe well the electronic structure of strongly correlated materials.
This rises the question whether the EPC 
is properly treated by the standard DFT-LDA approach for the 
``other high temperature superconductors"\cite{pickett2, Pickett} 
and other correlated materials in general.
Despite the answer to this question, a reliable evaluation of
electron-phonon interaction (EPI) strength for real materials is of fundamental importance
for both understanding the underlying physics and designing novel
functional materials. (The EPC for metallic materials 
is the average EPI matrix elements on the Fermi surface.)
The EPI/EPC plays an important role not only in conventional
superconductivity, but also in electronic 
transport, electronic-heat capacity, etc. 
Even in the cuprate superconductors, there are 
evidences\cite{LBCO, LMCO, HTSC1, HTSC2} 
(see also Ref.\onlinecite{cuprate-review} for a review) that EPC is strong and 
may play an important role in the 
unconventional superconductivity\cite{HTSC1}. 

The EPC for materials can be estimated by both experimental techniques and theoretical methods.
Experimentally, its strength can be 
estimated by means of neutron scattering\cite{Neutron}, 
Raman scattering\cite{Raman}, tunneling\cite{Tunneling2} and 
photoemission\cite{HTSC1}. 
However these experiments 
are usually limited to be able to estimate the EPC strength for only a few phonon 
modes.\cite{HTSC1, Raman, ARPES-C60, Neutron-C60} 
Theoretically, first principle calculations based on density functional theory
and linear response technique, 
on the other hand, can provide a complete evaluation of the EPI 
including all phonon modes, 
and is widely used for calculating lattice dynamical 
properties and EPC of solids. 
For example, the EPC evaluated by this approach 
with LDA/GGA are 
strong enough to account for the
rather high temperature superconductivity in \textit{conventional} superconductors such as elemental lithium
(T$_c\sim$20 K)\cite{deepa}, yttrium (T$_c\sim$20
K)\cite{yttrium}, calcium (T$_c\sim$25 K)\cite{calcium} and
binary compound MgB$_2$ (T$_c\sim$40 K)\cite{mgb2}.
This recorded success of LDA linear response calculations for conventional superconductors 
gives rise to the puzzles about the 
 mechanism of superconductivity in the ``other high temperature superconductors" 
including the bismuthates and the transition metal chloronitrides as mentioned above.

In this paper, we show that these ``other high temperature superconductors" 
are strongly coupled superconductors 
where the coupling of the electrons to the lattice vibrations are strongly enhanced 
by correlation effects which requires treatments of electron-electron interaction beyond LDA/GGA.
Once the EPC is properly evaluated, it is strong enough to account for their 
high temperature superconductivity.
Our view is supported by various experiments which show strong 
EPC for certain phonon modes of these superconductors.\cite{tunneling, linewidths, IXS, softening}

The key is to overcome the overscreening problem of LDA/GGA and achieve a simultaneous 
faithful description of the ground state electronic structures and lattice dynamical properties.
We will show that an appropriate screened hybrid functional and quasi-particle GW (QPGW) method provide 
an accurate description of the normal 
state electronic structures, and can be combined with linear 
response approach to accurately determine lattice dynamical properties and 
the EPC of these materials.
The problem is that the GW method and the screened hybrid functional approach are very computationally 
demanding, combining them directly to linear response approach pose a serious 
challenge to nowadays technology. It is therefore desirable to have a 
simplified method to estimate the EPC strength to 
a reasonable accuracy. 

To this end, we propose such a method based on LDA linear 
response calculations and a few GW and/or screened hybrid functional 
supercell calculations. The latter are used to replace the most important 
electron-phonon matrix elements in the LDA linear response calculations to 
obtain an improved evaluation of the EPC. 
We apply our method to the (Ba,K)BiO$_3$ and electron-doped HfNCl high temperature superconductors 
and find that the EPC computed by our method are 
significantly enhanced compared to the values predicted by LDA linear 
response calculations and are strong enough to account for the experimental 
T$_c$ in these materials within the standard Migdal-Eliashberg theory. 
Therefore the puzzling high temperature superconductivity in this 
class of superconductors is naturally explained. In addition, our method
explains the material and doping dependence of T$_c$ in these superconductors 
and related compounds. Our method can also be used to design new high temperature 
superconductors and other EPI/EPC related functional materials.

This paper is organized as following: we first clarify two sources of 
electronic correlations starting from LDA and their impacts on lattice 
dynamic properties (section II). 
A general understanding of the 
EPC and our method to estimate the EPC 
combining LDA linear response calculations and GW/screened 
hybrid supercell calculations is presented in section III. Section IV 
summarizes our computational details. The electronic structures, 
selected important phonon frequencies and EPC 
calculated by LDA functional, the screened hybrid functional, 
and QPGW approach are presented for 
(Ba,K)BiO$_3$ materials in Section V and electron doped HfNCl/ZrNCl 
materials in section VI. A short summary is provided in Section VII followed by discussions. 
The paper is concluded in Section VIII.

\section{Electronic correlation and its impact on lattice dynamic properties}

The frequency and momentum resolved green's function $G(\omega, k)$ can be written 
as $G(\omega, k)=1/(\omega-H(k)_{mean field}-\Sigma(\omega, k))$ where $H(k)_{mean field}$ 
is some mean field Hamiltonian and $\Sigma(\omega, k)$ is the self-energy correction 
induced by correlation effects. Naturally $\Sigma(\omega, k)$ depends on the choice of the mean field method, 
thus the term ``\textit{correlation}''. Chemists usually use Hartree-Fock method as the mean field whereas 
many physicists prefer LDA/GGA within the DFT framework as the starting point. Here we adopt the physicists' choice, i.e., 
$H(k)_{mean field}=H(k)_{DFT-LDA/GGA}$, and define correlation with respect to deviations from DFT-LDA/GGA. With this choice, 
the degree of correlation is quantified by $\Sigma(\omega, k)$. 

When $\Sigma(\omega, k)$ is negligible, the correlation is weak and static mean field theory such as 
DFT-LDA/GGA is reasonably accurate, for examples, Cu and Au. 
However, when $\Sigma(\omega, k)$ is large, correlation is strong and has to be treated by approaches beyond DFT-LDA/GGA. 
There are different cases where the correlation can be taken into account by different methods. 

If $\Sigma(\omega, k)$ is dominated by the frequency dependence, the correlation is mainly local and 
is caused by the degeneracy error of LDA/GGA.\cite{Yang} 
This local correlation is addressed by dynamical mean field theory (DMFT) 
combined with DFT-LDA/GGA.\cite{DMFT-RMP1996, DMFT-RMP2006}

On the other hand, if $\Sigma(\omega, k)$ has mostly the momentum dependence, the correlation is mostly 
non-local and is rooted in the semi-local nature of LDA/GGA which neglects the long-range exchange interaction. 
This non-local correlation can be accounted for by including long-range exchange interactions.
GW method is able to determine the range of the exchange potential self-consistently and is 
suited to deal with the non-local correlation. The disadvantage is that GW calculations are often 
very computationally demanding. The most economical way
to incorporate the long-range exchange interaction (non-local correlation) 
is by replacing the LDA/GGA functional by a
\textit{screened} hybrid functional\cite{HSE-review}, such as HSE06\cite{HSE2006},
which increases the spatial non-locality of the exchange potential relative to the LDA/GGA.
The downside of the screened hybrid functional approach
is that it contains an empirical mixing-parameter $\alpha$
(0.25 for HSE06) of the Hartree-Fock potential
and an screening parameter $\mu$ (0.2 for HSE06) to adjust
the range of the spatial non-locality and to avoid the spin-density wave
instability in the Hartree-Fock method.\cite{Overhauser, Hybrid-review1}
This drawback has already been faced a lot of criticism of the less
\textit{ab initio} nature compared to the LDA/GGA functional.
To avoid this, we fixed $\alpha=0.25$ for
all the screened hybrid functional calculations and we compare
the screened hybrid functional results with quasi-particle GW results
to determine the $\mu$ parameter.
Therefore, we have no free parameter in our screened
hybrid functional calculations and our approach preserves the
most valuable predictive power of \textit{ab initio} method.
It was shown that HSE06 share the same feature as QPGW in
accounting for the electronic structures and lattice dynamical
properties of the parent BaBiO$_3$ compound.\cite{hybrid1, hybrid2}

Finally, if both frequency and momentum dependence are important in $\Sigma(\omega, k)$, there are strong local 
and non-local correlations. To treat the local and non-local correlations simultaneously,  
we need to combine the above methods such as GW+DMFT [\onlinecite{GWDMFT1, GWDMFT2}] and screened hybrid DFT+DMFT[\onlinecite{HFDMFT}].

It has been established that the local correlation 
as seen in many partially filled $d$- and $f$-shell
materials often have strong effects on lattice dynamical properties. For
example, DMFT greatly improves LDA on the phonon spectra in Pu\cite{Pu-Savrasov}
and in the transition metal oxides MnO and NiO\cite{NiO-Savrasov}.
In the doped cuprate superconductors, the experimental width of the
half-breathing phonon is about an order of magnitude larger than predicted
by LDA calculations.\cite{phonon-Cuprates}

In parallel, the screened hybrid functional has been shown to 
improve over LDA/GGA the phonon spectra of many materials including 
semiconductors\cite{hybrid-semi1, hybrid-semi2, hybrid-semi3}, 
oxides\cite{hybrid-oxides}, and molecules\cite{hybrid-mo}.  
Note that the GW method and the (screened)
hybrid functional are usually used to correct the underestimated LDA/GGA band
gaps of some insulators and semiconductors. These approaches are generally
believed to have little effects on or perform worse than LDA for metallic
systems.\cite{Hybrid-review1, Hybrid-review2}
The impact of non-local correlation on the lattice dynamical properties and
EPC of metallic materials has not been carefully studied.

In this paper, we will show that the non-local correlation
can also have substantial impacts on the lattice dynamical properties of 
\textit{metallic} materials where LDA/GGA strongly overscreens the electronic states, 
such as conducting materials in the vicinity of a metal-insulator transition (i.e.,
the parent compound is an insulator, while the doped compound become
metallic upon sufficient doping).
We find that the \textit{screened} hybrid functional approach, when adjusted 
to match with QPGW results, is
able to reproduce the experimental electronic structures and phonon
frequencies of both the ``other high temperature superconductors"-the bismuthates and
the transition metal chloronitrides 
for which LDA/GGA calculations show large
discrepancies with experimental observations
and a conventional high temperature superconductor MgB$_2$ 
for which LDA calculations show less discrepancies with experimental observations.

\section{evaluation of electron-phonon coupling}
We now present our approach to evaluate the electron-phonon coupling.
The EPC strength $\lambda$ can by written as:\cite{Allen1} (the band index is omitted for simplicity)
\begin{equation}
\lambda=\frac{2}{N(\varepsilon_F)N_q}\sum_{k, q, \nu}|M_{k, k+q}^{\nu}|^2\delta(\varepsilon_k)\delta(\varepsilon_{k+q})/\omega_{q\nu}
\label{lambda}
\end{equation}
where N$_q$ is the number of $q$ points,
$N(\varepsilon_F)$ is the total density of states per spin at the Fermi level $\varepsilon_F$,
$\omega_{q,\nu}$ is the phonon frequency of branch $\nu$ with wave vector $q$,
and $M_{k, k+q}^{\nu}$ are the electron-phonon matrix elements given by
\begin{equation}
M_{k, k+q}^{\nu}=\sum_j (\frac{\hbar^2}{2M_j\omega_{q\nu}})^{1/2} \epsilon_{q,j}^{\nu}\cdot \langle k+q|\delta V/\delta u_{q, j}^{\nu}|k\rangle \\
\label{epm}
\end{equation}
where $j$ runs over the atoms in the unit cell and $\delta
V/\delta u_{q, j}^{\nu}$ is the partial derivative of the total
Kohn-Sham potential energy with respect to a given phonon
displacement $u_q^{\nu}$ of the $j$-th atom. For optical
vibration modes, with a wave vector $q$ commensurate with the lattice,
the electron-phonon matrix element
can be inferred from the shifts of the energy bands in a
supercell calculation (frozen phonons),
which we denote as reduced electron-phonon matrix element $D_{k, q}^{\nu}$ (REPME).
For states on the Fermi surface, the electron-phonon matrix elements can be
read directly from the splitting of the energy bands resulting
from the phonon displacements, which are shown in Fig.~\ref{band}.

As we shall show later, the modifications in more advanced approaches of the
Fermi surface, the density of states $N(\varepsilon)$ and the
corresponding phonon frequencies $\omega_{q, \nu}$ compared to LDA
are significantly smaller than those of the electron-phonon matrix elements. 
Hence, the electron-phonon matrix elements is the dominating factor
differing the realistic $\lambda$ from the LDA value.
As a result the realistic EPC $\lambda$
can be estimated from a LRT-LDA calculation by
rescaling the LDA REPMEs to the actual values given by more advanced approaches 
while keeping the integral over $k$ and $q$ of Eq.~\ref{lambda} at the LDA level.
Here the more advanced approach should be able to accurately reproduce the ground state properties (comparing to LDA).
In practice, a good estimation is achieved by using both an advanced approach and LDA to evaluate
the REPMEs for all the strongly coupled phonon modes (which can be seen in the LDA linear response calculation)
at special points in the Brillouin zone:
\begin{equation}
\lambda_A = \sum_{\nu} \lambda_{A\nu}\simeq \sum_{\nu}
\lambda_{L\nu} \langle |D_A^{\nu}|^2 \rangle/\langle |D_L^{\nu}|^2\rangle
\label{lratio1}
\end{equation}
where we denote the advanced approach (such as GW and screened hybrid functional approaches) as A and LDA as L.
If the enhancements in the REPMEs of all the strongly coupled branches
are of comparable magnitude we can estimate $\lambda_A$ by
\begin{equation}
\lambda_A \simeq \lambda_L\langle |D_A^{\nu}|^2 \rangle/\langle |D_L^{\nu}|^2 \rangle
\label{lratio2}
\end{equation}

We note that the accuracy of estimating the realistic EPC using the above method can be improved by including more phonon modes 
when more computing resource is available. Eventually, to achieve a more accurate evaluation of the realistic EPC, linear response technique should 
be implemented into GW and/or screened hybrid functional DFT.   

\section{Computational details}
\subsection{Method}
The electronic structures are calculated using
both the VASP[\onlinecite{vasp}] code with GGA(PBE)[\onlinecite{PBE}] and HSE06[\onlinecite{HSE2006}] exchange-correlation functionals and WIEN2k[\onlinecite{wien2k}] code
with GGA(PBE)[\onlinecite{PBE}] exchange-correlation functional. 
Linear response calculations are carried out using the LMTART[\onlinecite{lmtart1, lmtart2}] code 
with the ideal cubic structures (i.e., without distortions). 
The crystal structures are taken from Ref.\onlinecite{struct-BBO} for BaBiO$_3$ and Ref. \onlinecite{struct-BPO} for BaPbO$_3$.
For VASP calculations, the PAW-PBE pseudopotential is used in both DFT-GGA and DFT-HSE calculations.
The energy cutoff for the wavefunction is 400 eV.
The results of GGA calculations from VASP are double checked with results from WIEN2k with GGA(PBE) exchange-correlation functional.
In linear response calculations, a 32$^3$ k mesh is used to converge the electron charge density.
The electron phonon coupling constant is obtained with a 8$^3$ q mesh with 35 independent q points in the irreducible Brillouin zone.
We use virtual crystal approximation to simulate the doping effect in the linear response calculations.

For GW calculations, we use the self-consistent quasi-particle GW (scQPGW) as implemented in VASP. 
To ensure good convergence of the scQPGW calculations, we use an energy cutoff of 400 meV, 
480 bands and 6$\times$6$\times$6 k-mesh for the 10-atom/cell (Ba,K)BiO$_3$ and 
240 bands and 12$\times$12$\times$4 k-mesh for the 6-atom/cell Hf/ZrNCl compounds. 

\subsection{Frozen phonon calculations}
For the frozen-phonon calculations, we create supercells adaptive to the momentum $Q$ of the phonon mode.
For example, to calculate the oxygen breathing frequency at $X$, $M$ and $R$ points in bismuthates,
we create 2$\times$1$\times$1, 2$\times$2$\times$1, and
2$\times$2$\times$2 supercells of the simple cubic unit cell, respectively.
Upon atomic displacement $u$ of a phonon mode with wavevector $Q$,
the phonon frequency is extracted as $\omega=(E''(u)|_{u=0}/M_{eff})^{1/2}$ in the harmonic approximation, where
$M_{eff}$ is the effective mass and $E(u)$ is the total energy as a function of the atomic displacement $u$
in the DFT frozen phonon calculation.
For instance, the effective mass $M_{eff}$ is 6 times of oxygen atomic mass in calculating the oxygen breathing frequency at $R$ point.

\subsection{Estimation of $\omega_{log}$}

While $\omega_{log}$ usually decreases with increasing $\lambda$, there isn't a simple relation between them.
Following a discussion in Ref.~[\onlinecite{softening}],
we estimate the HSE06 value $\omega_{log, H}$ in optimal hole-doped BaBiO$_3$ (and other compounds in Table I of the main article) from the
corresponding LDA value $\omega_{log, L}\simeq$550 K[\onlinecite{Meregalli}]
via an empirical relation as $\omega_{log, H}\simeq\omega_{log, L} (1+\lambda_L)^{1/2}/(1+\lambda_H)^{1/2}\simeq$450 K which
is in good agreement with the value about 450 K extracted from experiments[\onlinecite{tunneling, PhononDOS}].
The good agreement of the $\omega_{log}$ in optimal hole-doped BaBiO$_3$ between the estimated HSE06 value and experimental value suggests,
a) the above relation is a reasonable approximation and b) the HSE06 $\lambda_H$ is close to the realistic EPC in this material which
further supports our view that HSE06 is appropriate for estimating the EPC in BaBiO$_3$-related materials.

\subsection{Evaluating Coulomb pseudopotential}

To estimate the Coulomb pseudopotential $\mu^*$, we implemented
the methodology of Ref.[\onlinecite{KHLee}] in our
codes[\onlinecite{jcm_15_2607,prb_80_041103}] which combine LDA and GW
methods in a linearized augmented plane-wave (LAPW) basis set.
$\mu^{*}$ is given by the formula
\begin{align}\label{mu_1}
\mu^{*}=\frac{\mu}{1+\mu\ln\frac{\varepsilon_F}{\omega_D}} 
\end{align}
where $\varepsilon_F$ is the Fermi energy and $\omega_D$ the Debye energy,
and dimensionless parameter $\mu$ is calculated as the following
average over Fermi surface
\begin{align}\label{mu_2}
\mu=N(E_{F})\frac{\sum_{\mathbf{k}\mathbf{k}'}\sum_{\lambda\lambda'}W_{\mathbf{k}\lambda;\mathbf{k}'\lambda'}
\delta(\varepsilon_{\mathbf{k}\lambda}-E_{F})\delta(\varepsilon_{\mathbf{k}'\lambda'}-E_{F})}{\sum_{\mathbf{k}\mathbf{k}'}\sum_{\lambda\lambda'}
\delta(\varepsilon_{\mathbf{k}\lambda}-E_{F})\delta(\varepsilon_{\mathbf{k}'\lambda'}-E_{F})},
\end{align}
where $N(E_{F})$ is the density of states at Fermi level per spin,
$\varepsilon_{\mathbf{k}\lambda}$ is the Kohn-Sham eigenvalue of
the $\lambda$th band at a wave vector $\mathbf{k}$, and matrix
elements of the screened Coulomb interaction in a basis of LDA
eigenstates are calculated as the following
\begin{align}\label{mu_3}
W_{\mathbf{k}\lambda;\mathbf{k}'\lambda'}=\langle\mathbf{k}'\lambda'\uparrow,-\mathbf{k}'\lambda'\downarrow|W|
\mathbf{k}\lambda\uparrow,-\mathbf{k}\lambda\downarrow\rangle.
\end{align}
The resulting $\mu^*$ for the ideal BaBiO$_3$, BaPbO$_3$ and LaPbO$_3$
compounds are 0.104, 0.117 and 0.119 respectively.

The difference between our work and Ref. \onlinecite{KHLee}
is that we use finite temperature formalism and approximate $\delta$
function with the imaginary part of LDA Matsubara's Green's function
\begin{align}\label{mu_4}
\delta(\varepsilon_{\mathbf{k}\lambda}-E_{F})\approx Im
\frac{1}{i\omega_{0}+E_{F}-\varepsilon_{\mathbf{k}\lambda}},
\end{align}
where $\omega_{0}$ is the smallest positive Matsubara's frequency.
The expression (\ref{mu_4}) becomes exact in the limit of zero
temperature.
Technically we perform the following steps. First we calculate LDA
one-electron spectrum and corresponding wave functions. It provides
us with LDA Green's function. Then we evaluate LDA polarizability
$P_{LDA}=G_{LDA}G_{LDA}$. The third step is to calculate screened
interaction $W=V+VP_{LDA}W$, where $V$ is bare Coulomb interaction.
After the third step we are able to get the Coulomb pseudopotential
parameter.

\section{Bismuthates}

We first apply our method to the bismuthates Ba$_{1-x}$K$_x$BiO$_3$ which was 
found to superconduct below 32 K in 1988.\cite{Cava} 

\subsection{Crystal Structure}

\begin{figure}[htb]
\includegraphics[width=0.99\linewidth]{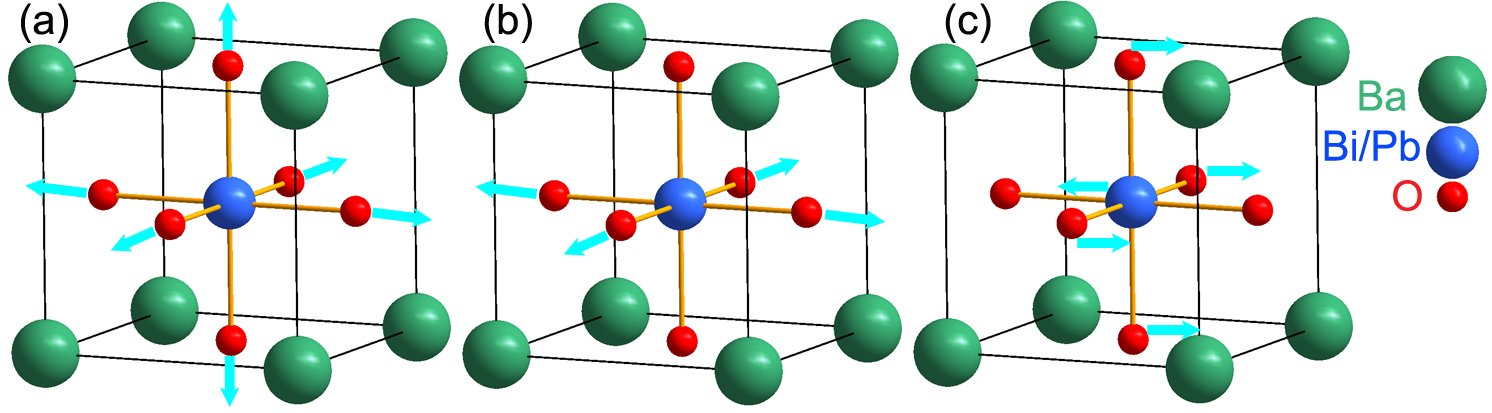}
\includegraphics[width=0.99\linewidth]{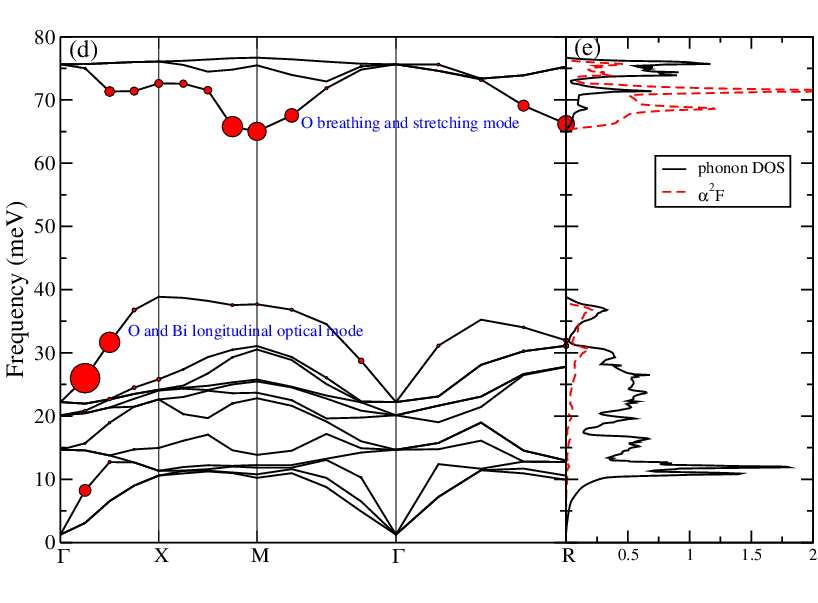}
\caption{
(Color online)
Top panel: Crystal structure of BaBiO$_3$ and BaPbO$_3$.
The arrows show (a) the oxygen breathing, (b) the oxygen stretching
and (c) an oxygen and bismuth longitudinal optical vibration modes (the ferroelectric longitudinal optical mode).
Bottom panel: (d) (left) phonon spectra and mode- and momentum $q$-dependent electron-phonon
coupling strength $\lambda_{q, \nu}$ (its value is proportional
to the size of the circle at each $q$ point)
and (e) (right) the corresponding phonon density of states and
the Eliashberg functional $\alpha^2F(\omega)$ of 0.4 hole doped BaBiO$_3$.
}
\label{struct}
\end{figure}

The optimal doped BaBiO$_3$ (and BaPbO$_3$)
crystallize in the simple cubic
perovskite structure as shown in Fig.\ref{struct},
whereas the parent compounds (without doping) display
certain distortions from the ideal perovskite structure.\cite{struct-BBO, struct-BPO}
There are two types of distortions:
one is the oxygen breathing distortion along nearest-neighbor $M$-O bonds ($M$=Bi and Pb) as shown in Fig. \ref{struct}(a).
The other one is the oxygen tilting distortion perpendicular to the nearest-neighbor $M$-O bonds (not shown).
The oxygen breathing vibration was found to show strong EPC.\cite{Mattheiss, Shirai, Meregalli}

\subsection{LDA linear response calculation}

We first reproduce the LRT-LDA calculations of Ref.
\onlinecite{Meregalli} and show the results in Fig.~\ref{struct}(d) and (e).
We find that the oxygen breathing/stretching mode around $R$ point
(Fig.\ref{struct}(a)) and around $M$ point
(Fig.\ref{struct}(b)) dominate the contributions to the total
$\lambda$, as evident from the mode- and momentum- dependent EPC
$\lambda_{q, \nu}$ shown in Fig.\ref{struct}(d) and the
Eliashberg function $\alpha^2F(\omega)$ shown in
Fig.\ref{struct}(e). In addition, we find at small $q$ a
longitudinal optical mode (so-called ferroelectric mode)
depicted in Fig.\ref{struct}(c) which
also has substantial EPC. These findings are in good agreement
with previous observations as well.\cite{Mattheiss, Shirai,
Meregalli} On the contrary, acoustic phonon modes have very little
contribution to the total EPC. With these findings, we argue that
it is appropriate to use equation (\ref{lratio1}) to estimate the
realistic EPC strength $\lambda_A$ from LRT-LDA calculations in
(Ba,K)BiO$_3$.

\begin{figure}[htb]
\includegraphics[width=0.9\linewidth]{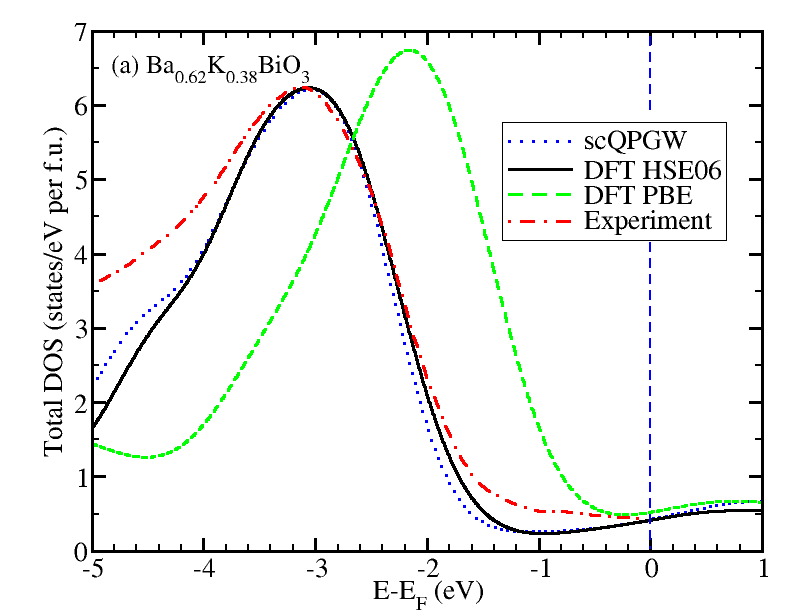}
\includegraphics[width=0.9\linewidth]{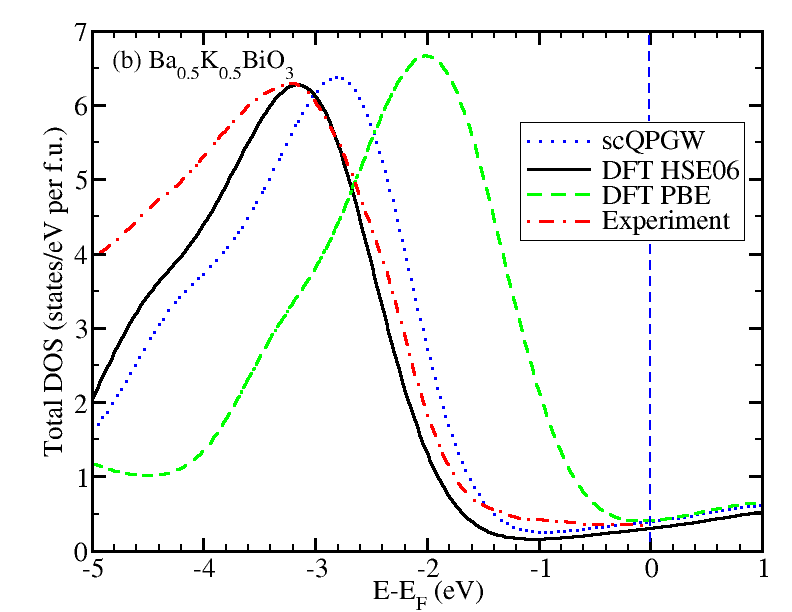}
\caption{
\textbf{Density of states of Ba$_{0.62}$K$_{0.38}$BiO$_3$ and Ba$_{0.5}$K$_{0.5}$BiO$_3$.}
Top panel: the DOS of Ba$_{0.62}$K$_{0.38}$BiO$_3$ calculated using
the PBE functional, the HSE06 screened hybrid functional and the self-consistent quasiparticle GW,  where doping is approximated
with the rigid band approximation.
Bottom panel: the DOS of Ba$_{0.5}$K$_{0.5}$BiO$_3$ calculated using the PBE functional, the HSE06 screened hybrid functional and 
the self-consistent quasiparticle GW.  
Here the doping is simulated with a 2x2x2 supercell by substituting half of the Ba atoms with K atoms.
The experimental data is the ultraviolet photoemission spectra for (a) Ba$_{0.62}$K$_{0.38}$BiO$_3$
and (b) Ba$_{0.52}$K$_{0.48}$BiO$_3$ taken from Ref. \onlinecite{BBO-PES}. }
\label{DOS}
\end{figure}

\subsection{Density of States}

It was noted\cite{Meregalli} that LDA suffers large discrepancies in describing the ground state properties 
of the parent compounds and some lattice dynamical properties of the metallic doped compounds in comparison with experiments. 
Recently, it was shown that a screened hybrid functional such as HSE06\cite{HSE2006} 
largely removes the overscreening of LDA/GGA
in the insulating (parent and lightly hole-doped) BaBiO$_3$ compounds and describes many of 
their physical properties in excellent agreement with experiments.\cite{hybrid1, hybrid2} 
It is however not clear if the screened hybrid functional also provides an improved description of the 
metallic hole-doped BaBiO$_3$ compounds.

Here we show in Fig.\ref{DOS}(a) the calculated density of states (DOS) of
an optimal doped compound Ba$_{0.62}$K$_{0.38}$BiO$_3$ using both the
PBE functional and HSE06 functional in the DFT framework, and self-consistent quasiparticle GW method, 
based on virtual crystal approximation (VCA) and rigid band
approximation (RBA), and we compare them with the experimental
ultraviolet photoemission spectra[\onlinecite{BBO-PES}] (UPS) of Ba$_{0.62}$K$_{0.38}$BiO$_3$.
We note that the difference between RBA and the VCA is small with the PBE functional 
therefore only the RBA result is displayed in Fig.\ref{DOS}(a). 
The HSE06 DOS is almost the same as the scQPGW DOS, and both of them have an overall good agreement
with the experimental UPS, in
particular they share roughly the same peak of the oxygen $2p$ states centered at about -3.1 eV. 
In contrast, the DOS from PBE calculation has a peak
centered at about -2.2 eV, 0.9 eV away from the experimental position.

In Fig.\ref{DOS}(b) we show the DOS for Ba$_{0.5}$K$_{0.5}$BiO$_3$ calculated with PBE and HSE06 functionals, and scQPGW method.
Instead of using VCA or RBA, we use a 2$\times$2$\times$2 supercell and replace half of the Ba atoms with K atoms to minimize
errors induced by such approximations. We compare the results with the experimental UPS data for Ba$_{0.52}$K$_{0.48}$BiO$_3$
from Ref.\onlinecite{BBO-PES}. Again we find very good agreement between the HSE06 calculated DOS and experimental spectral. 
The peak position of the scQPGW DOS is 0.4 eV off the experimental one whereas the peak position of the PBE DOS is 1.2 eV off the experimental one.

Therefore, both the scQPGW method and HSE06 \textit{screened} hybrid functional provide a significantly improved 
description of the electronic structures of the \textit{metallic} (Ba,K)BiO$_3$ compounds compared to the LDA results, 
in strong contrast to the common belief that GW and hybrid functional are worse than LDA for metals.

\begin{table*}[htb]
\caption{
Phonon frequencies of the important longitudinal optical phonon modes at high symmetry points in BaKBi$_2$O$_6$
calculated by supercell frozen phonon calculations in the DFT-GGA and DFT-HSE06 framework.
For comparison, the corresponding phonon frequencies from LRT-LDA calculations
for Ba$_{0.6}$K$_{0.4}$BiO$_3$ and Ba$_{0.5}$K$_{0.5}$BiO$_3$ using the virtual crystal approximation (VCA),
as well as from experimental measurements (exp.1 from Ref.\onlinecite{BKBO-phonon} and exp.2 from Ref.\onlinecite{softening}) are also displayed.
 }

\label{Phonon-frequency}
\begin{tabular}{|c|c|c|c|c|c|c|c|c|c|c|}
\hline
          & $Q$        & DFT-GGA   & DFT-HSE06  & DFT-GGA & LRT-LDA  & LRT-LDA &  exp.1 & exp.2  \\
Doping method &  & K substitution  & K substitution & VCA & VCA      & VCA     & K substitution   &  K substitution      \\
\hline
Mode/Doping $x$ &      & 0.5       & 0.5        & 0.5         & 0.5      & 0.4     & 0.4    & 0.52  \\
\hline
Breathing & $R$        & 68.4      & {\bf 61.2}       & 72.3        & 75.1     & 68.6    & {\bf 62}     &       \\
          & $M$        & 64.7      & {\bf 59.0}       & 64.3        & 68.8     & 67.9    & {\bf 60}     &       \\
          & $X$        & 66.6      & {\bf 60.1}       & 71.0        & 74.2     & 75.9    &{\bf 55}     & {\bf 55}    \\
Ferroelectric & $\Gamma$ & 22.7    & {\bf 18.1}       & 22.9        & 23.8     & 23.6    & 25     & {\bf 18}    \\
          & $X$        & 37.2      & {\bf 38.1}       & 38.6        & 39.7     & 39.7    & {\bf 36}     & {\bf 35}     \\
\hline
\end{tabular}
\end{table*}

\subsection{Phonon frequencies}

It is not feasible to compute the phonon frequencies using the scQPGW method currently. 
Here we provide evidence that HSE06 also improves the
estimation of phonon-related quantities over LDA/GGA for the metallic (Ba,K)BiO$_3$ compounds. 
The scQPGW phonon frequencies are expected to be similar to the HSE06 results.
We show in Table \ref{Phonon-frequency}
the phonon frequencies of the oxygen breathing/stretching mode at $X$, $M$, $R$ points and
the so-called ferroelectric longitudinal optical phonon mode (see Fig.\ref{struct})
at $\Gamma$ and $X$ points in BaKBi$_2$O$_6$ (a realistic K doping
into BaBiO$_3$) computed using the frozen-phonon method with both GGA and HSE06.
We display also in Table \ref{Phonon-frequency} the phonon frequencies
computed by DFT-GGA and LRT-LDA with the virtual crystal approximation (VCA)
for Ba$_{0.6}$K$_{0.4}$BiO$_3$ and Ba$_{0.5}$K$_{0.5}$BiO$_3$
and compared with available experiments from Ref.\onlinecite{BKBO-phonon} for Ba$_{0.6}$K$_{0.4}$BiO$_3$
and Ref.\onlinecite{softening} for Ba$_{0.48}$K$_{0.52}$BiO$_3$.
For the oxygen breathing mode at $M$ and $R$ points, we find the HSE06 frequencies are in very good agreements
with experimental data whereas GGA overestimates the experimental frequencies by about $10\%$.
The HSE06 improves GGA on the oxygen breathing frequency at $X$ point, where
HSE06(GGA) overestimates the experimental value by $9\%$($21\%$).
For the ferroelectric longitudinal optical mode, the phonon frequency at $X$ point is similar
in HSE06 and GGA as expected from the small $\lambda_{q, \nu}$ shown in Fig.\ref{struct}.
However, at $\Gamma$ point, the HSE06 phonon frequency of this mode (18.1 meV)
is considerably smaller than the GGA value (22.7 meV) due to the
enhanced $\lambda_{q, \nu}$ and is in good agreement with experimental observation ($\sim$18 meV)[\onlinecite{softening}].

\subsection{Band structures and reduced electron-phonon matrix elements}

\begin{figure}[htb]
\includegraphics[width=0.9\linewidth]{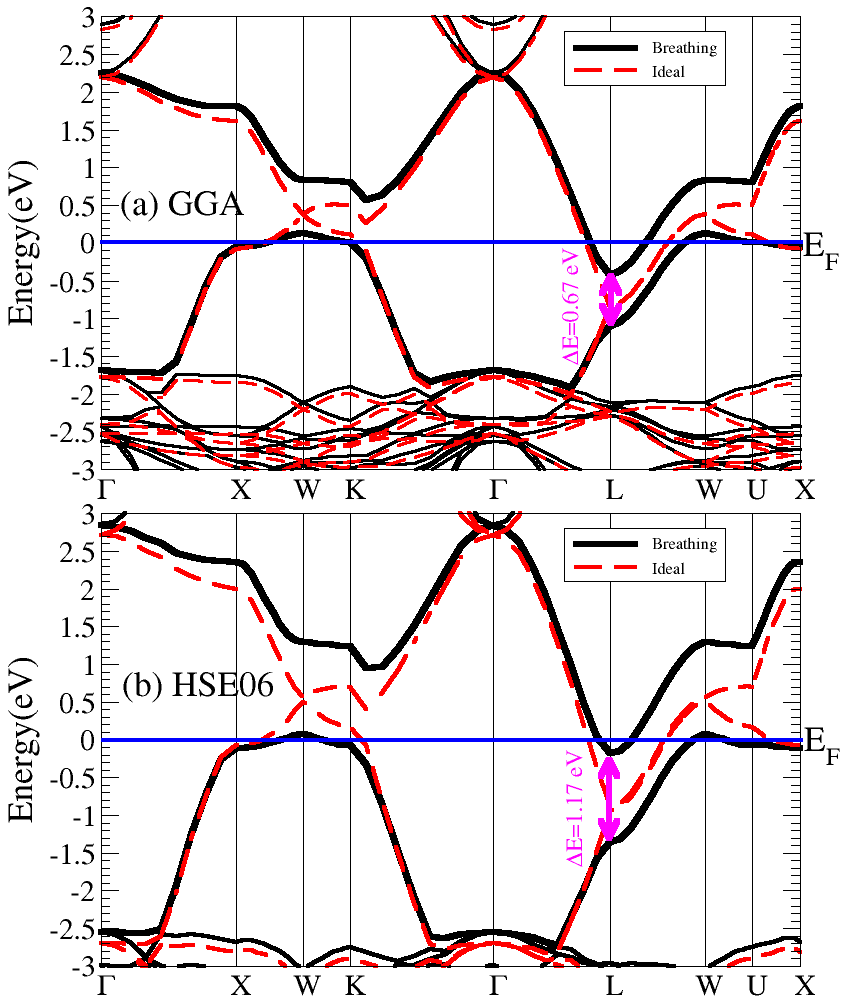}
\caption{
(Color online)
Illustration of the big REPME of the oxygen breathing mode at $R$ point in BaBiO$_3$.
The band structures of BaBiO$_3$ with and without oxygen breathing displacement calculated
by DFT using both the (a) GGA and (b) HSE06 hybrid functionals.
We plot the band structure in the Brillouin zone of the $fcc$ unit cell
corresponding to a 2$\times$2$\times$2 supercell of the simple cubic unit cell
due to the oxygen-breathing distortion.
The displacement of each oxygen is about 0.044 $\AA$ in the calculations.
The band splittings indicated by the arrows are about 1.17 eV for DFT-HSE06 and 0.67 eV
for DFT-GGA, resulting in REPMEs of about 13.3 and 7.6 eV/$\AA$, respectively.
The corresponding scQPGW values are 1.10 eV and 12.5 eV/$\AA$, respectively.
Note that the material remains metallic with the oxygen breathing distortion.
}
\label{band}
\end{figure}

\begin{figure}[htb]
\includegraphics[width=0.99\linewidth]{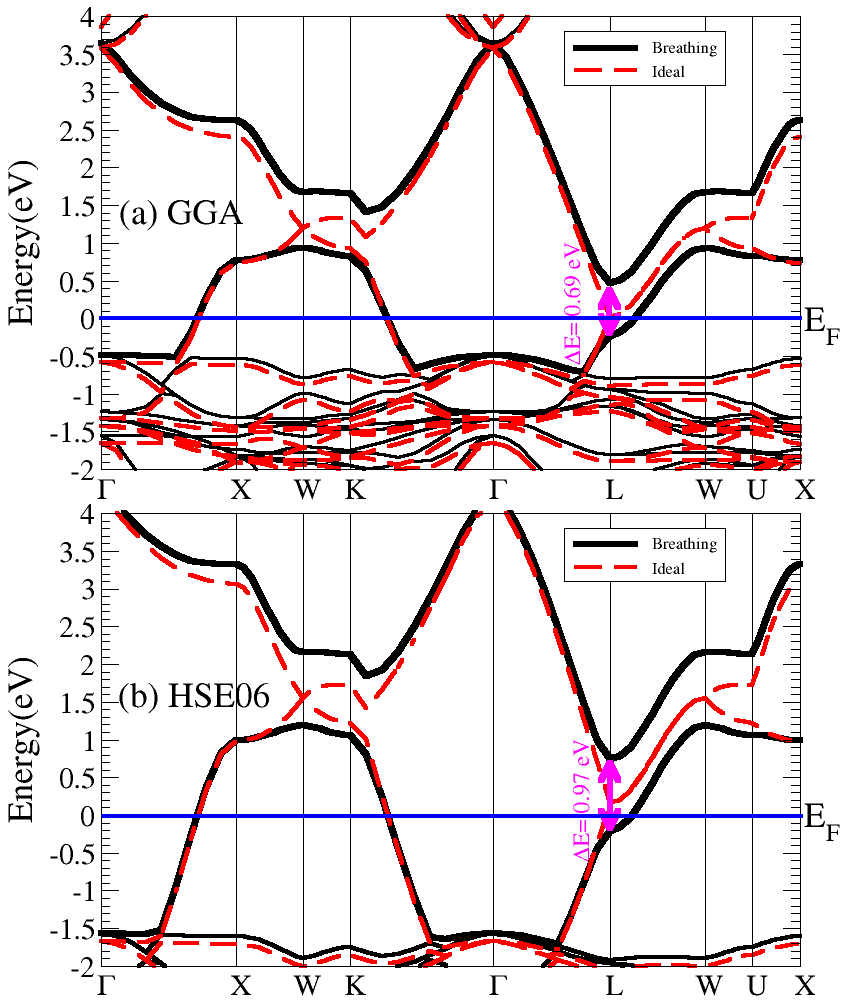}
\caption{
(Color online)
Illustration of the big REPME of the oxygen breathing mode at $R$ in BaKBi$_2$O$_6$.
The band structures of BaKBi$_2$O$_6$ with and without oxygen breathing displacement calculated
by DFT using both the (a) GGA and (b) HSE06 hybrid functionals.
We plot the band structure in the Brillouin zone of the $fcc$ unit cell
corresponding to a 2$\times$2$\times$2 supercell of the simple cubic unit cell
where half of the Ba atoms are replaced by K atoms.
The displacement of each oxygen is about 0.044 $\AA$ in the calculations.
The band splittings indicated by the arrows are about 0.97 eV for DFT-HSE06 and 0.69 eV
for DFT-GGA, resulting in REPMEs of about 11.0 and 7.8 eV/$\AA$, respectively.
The corresponding scQPGW values are 1.21 eV and 13.7 eV/$\AA$, respectively.
Note that the material remains metallic with the oxygen breathing distortion.
}

\label{band2}
\end{figure}

As an example, we show in Fig.\ref{band} the band structures
of BaBiO$_3$ with and without the oxygen breathing displacement (see Fig.~\ref{struct}(a))
to illustrate the huge differences in the REPME of the oxygen breathing
mode at $R$ point computed by GGA and HSE06.
The band splittings indicated by the arrows in Fig.\ref{band} are about 1.17 eV for DFT-HSE06 and 0.67 eV
for DFT-GGA, upon a displacement of 0.044 $\AA$, resulting in REPMEs of about 13.3 and 7.6 eV/$\AA$, respectively.
We also compute this splitting using scQPGW under the same displacement.
The corresponding value is 1.10 eV, resulting in a REPME of about 12.5 eV/$\AA$. 
The scQPGW results confirm again that HSE06 is applicable to the metallic (Ba,K)BiO$_3$ compounds.
The above results indicate that the EPC strength for this mode is strongly enhanced by about 
a factor of three in the more accurate HSE06 and scQPGW treatments compared to the LDA value.
Notice that the material remains metallic with the oxygen breathing distortion, therefore the enhanced REPME 
in the HSE06 and scQPGW approaches compared to the LDA value is not merely a band gap problem of LDA.

We also compute the REPME of the oxygen breathing mode at $R$ point (see Fig.\ref{band2}) in
BaKBi$_2$O$_6$ and find it to be about 13.7, 11.0, and 7.8 eV/$\AA$ in
scQPGW, HSE06, and GGA, respectively. 
Notice again that the material remains metallic with the oxygen breathing distortion.
Therefore, even in the overdoped
Ba$_{0.5}$K$_{0.5}$BiO$_3$ compound, the EPC is strongly enhanced
in scQPGW (HSE06), about a factor of three (two) of the LDA/GGA value, confirming
that the effects of doping on the REPME is relatively weak.
Hence, while the LDA/GGA quantitative overestimation of the phonon
frequencies is less severe than its underestimation of the REPME,
they are consistent manifestations of the overscreening problem of
LDA/GGA and calculations with the HSE06 functional
brings both quantities in closer agreement with experiments.

\subsection{Realistic electron-phonon coupling and T$_c$}

To estimate the realistic EPC, we compute the REPMEs 
of the three most important phonon modes as suggested by LDA linear response calculations.
These phonon modes are depicted in Fig.~\ref{struct}(a-c). 
The REPMEs of the oxygen breathing mode with Q vector corresponding to $R$ point (Fig.~\ref{struct}(a) have been 
discussed above, about 13.3, 12.5 and 7.6 eV/$\AA$ in HSE06, scQPGW and GGA calculations, respectively.
Since scQPGW has more or less the same results as HSE06, we use only HSE06 and GGA to calculate the REPMEs for 
the other two phonon modes (Fig.~\ref{struct}(b-c)), whose values
are about 8.9 and 5.7 eV/$\AA$ in HSE06 but only
5.1, and 3.4 eV/$\AA$ in GGA, respectively.
For all the three important phonon modes, we have $|D_H^{\nu}|^2/|D_L^{\nu}|^2\simeq$ 3.0.
Based on the LRT-LDA calculated EPC $\lambda_L\simeq0.33$,
we estimate the realistic electron phonon coupling $\lambda_H \simeq 3.0\lambda_L\simeq$1.0 for optimal hole-doped BaBiO$_3$.

At low doping, the EPC is strong enough to drive the
material to a polaronic state,\cite{hybrid1}
whereas for optimal doped BaBiO$_3$
(eg. Ba$_{0.6}$K$_{0.4}$BiO$_3$), we expect the Migdal-Eliashberg theory
is valid in this region.
Using the modified McMillan equation
\begin{equation}
T_c=\frac{\omega_{log}}{1.20}exp(-\frac{1.04(1+\lambda)}{\lambda-\mu^*(1+0.62\lambda)})
\label{eq3}
\end{equation}
with estimated $\omega_{log} \simeq 450$ K (see section III (C))
and $\lambda = 1.0$, $T_c$ is estimated to be 31 K (see
Table~\ref{deformation-and-Tc}) with $\mu^*$=0.1 which is the
conventionally nominated value and is consistent with the
value obtained by our first principles calculations as detailed in Section III (D).
As a result, the strong EPC strength $\lambda \simeq 1.0$ is
enough to explain the high
$T_c \simeq$ 32 K in K-doped BaBiO$_3$ in the framework of
a novel correlation-enhanced phonon mediated mechanism.

\subsection{Material and doping dependence of T$_c$}

\subsubsection{BaPbO$_3$}
We note that the previous reasoning can also be used to estimate
the EPC in one compound from another isostructural compound by
comparing their REPMEs, provided that they have similar band
structure around the Fermi level when optimally doped.

We also carried out LRT-LDA calculations in Ba$_{1-x}$La$_x$PbO$_3$
based on virtual crystal approximation. We find a strong doping
dependent $\lambda_L$, which is very small at small La doping and
reaches a maximum value of 0.58 at $x=0.7$. 
The EPC in optimal doped (Ba,La)PbO$_3$ is thus estimated to be about
0.58 (LDA) and 0.72 (HSE06). Our calculation suggests
superconductivity with T$_c$ up to 18 K in optimal doped
(Ba,La)PbO$_3$ compound.(see Table~\ref{deformation-and-Tc})
Experimentally 11 K superconductivity was found in a
Ba$_{1-x}$La$_x$PbO$_3$ multiphase compound synthesized at high
pressure, but the crystal structure of the superconducting phase
was not identified.\cite{BLPO} It would be very interesting to
synthesize high-quality single-crystal Ba$_{1-x}$La$_x$PbO$_3$
compound in the perovskite structure and test our theory.

\subsubsection{Ba$_{n+1}$Bi$_n$O$_{3n+1}$}
Our work sheds light on mystery of the dimensionality dependence
of T$_c$ in BaBiO$_3$-related materials. In heavy fermion materials,
low dimensionality enhances superconductivity relative to their three
dimensional versions.\cite{PMonthoux}
On the other hand in BaBiO$_3$-related materials,
layering, as in the synthesis of
Ba$_{n+1}$Bi$_n$O$_{3n+1}$ ($n$=1, 2, 3, $\cdots$) degrades T$_c$.
We evaluate the REPME
associated with O-breathing vibration and found it to be almost
zero for the Ba$_3$Bi$_2$O$_7$ compound ($n$=2 member), resulting
in very weak EPC and non-superconductivity, as seen in
experiments\cite{BBO327}. While general factors may promote
superconductivity in layered materials relative to their three
dimensional version, a system specific calculation for the
BaBiO$_3$ family reveals that in this case the dominant effect is the
reduction of the coupling of the phonon modes to the electrons
in the two dimensional materials and this results in a reduction
of T$_c$.

\section{transition metal chloronitrides}
We turn to another member of the ``other high temperature
superconductors": the electron-doped $M$N$X$ ($M$=Ti, Zr, and Hf;
N=nitrogen; $X$=Cl, Br, and I).  These materials are also
proximate to an insulating state and their T$_c$ are 16.5, 16 and
25.5 K for electron-doped $\alpha$-TiNCl, $\beta$-ZrNCl and
$\beta$-HfNCl, respectively.\cite{TiNCl, MNX} A LRT-LDA calculation
reported $\lambda$ about 0.52 and $\omega_{log}$ about 36.4 meV
in 1/6 electron-doped ZrNCl,\cite{ZrNCl} which gives T$_c$ only 6 K (assuming
$\mu^*$=0.1) and is insufficient to account for its 16 K
superconductivity, raising again the question of what is the
mechanism of superconductivity in this family.

\begin{figure}[htb]
\includegraphics[width=0.99\linewidth]{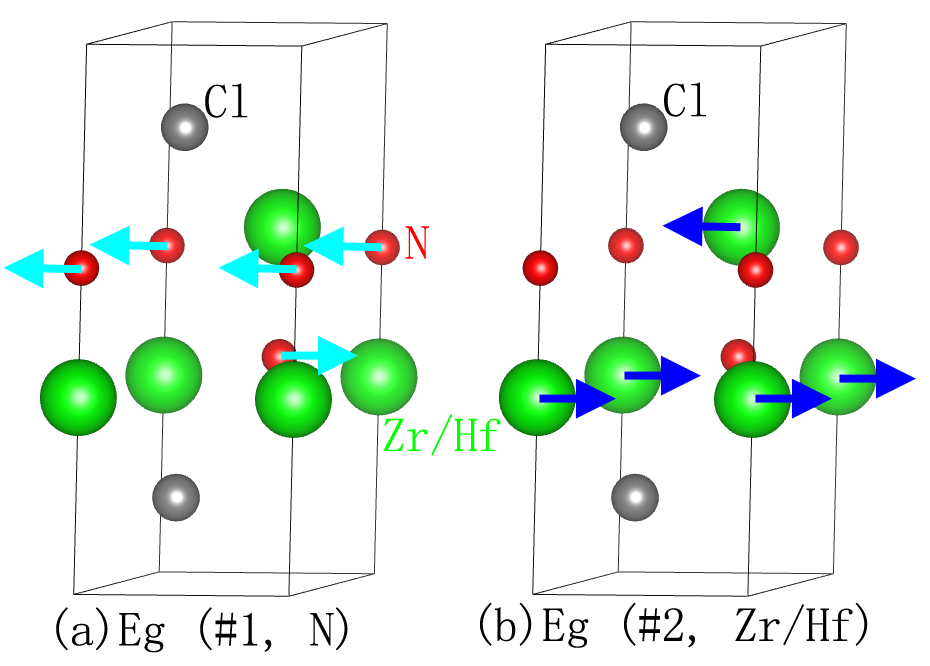}
\caption{
(Color online) Simplified crystal structure of ZrNCl/HfNCl.
The arrows show (a) the $E_g$ mode with in-plane N vibration, (b) $E_g$ mode with in-plane Zr/Hf vibration.
}
\label{struct-ZrNCl}
\end{figure}

\subsection{Crystal Structure}

ZrNCl and HfNCl crystallizes in a rhombohedral structure whose unit cell is built up with three
neutral (ZrNCl)$_2$ units in a hexagonal structure. The neutral (ZrNCl)$_2$ units are connected \textit{via}
weak van der Waals force along the $c-axis$. Since the adjacent layers of the (ZrNCl)$_2$ units are weakly coupled,
we ignore the shifting between the layers for simplicity, which results in a hexagonal unit cell (Space group $P\bar{3}m1$, \#164,
see Fig.\ref{struct-ZrNCl}).

In Fig.\ref{struct-ZrNCl}, we show two $E_g$ vibration modes involving mainly in-plane vibrations of the nitrogen atoms (Fig.\ref{struct-ZrNCl}(a))
and the Zr/Hf atoms (Fig.\ref{struct-ZrNCl}(b)). For simplicity, we ignore the small vibrations of the rest atoms in each mode.
Such a simplification has negligible influence on the calculated phonon frequencies as shall be shown in the following subsection.

We include in Table~\ref{ZrNCl-struct} the structural parameters used in our calculations for ZrNCl, Li$_{0.25}$ZrNCl, HfNCl and Na$_{0.25}$HfNCl.
Note that we simplify the structure to a single layer of the (Zr/HfNCl)$_2$ unit. This simplification is expected to have little impact on
the calculated electronic structures and lattice dynamics, due to the weak van der Waals force between the layers.

\begin{table}[htb]
\caption{
Crystal structures of ZrNCl, Li$_{0.25}$ZrNCl, HfNCl and Na$_{0.25}$HfNCl used in our calculations.
The atomic positions are Zr/Hf: (0, 0, $z(Zr/Hf)$) and (2/3, 1/3, -$z(Zr/Hf)$); N: (0, 0, $z(N)$) and
(2/3, 1/3, -$z(N)$), and Cl: (1/3, 2/3, $z(Cl)$) and (1/3, 2/3, -$z(Cl)$).
Note in the space group $P\bar{3}m1$, the atomic positions are shifted to
Zr/Hf: (2/3, 1/3, $z(Zr/Hf)$) and (1/3, 2/3, -$z(Zr/Hf)$); N: (2/3, 1/3, $z(N)$) and
(1/3, 2/3, -$z(N)$), and Cl: (0, 0, $z(Cl)$) and (0, 0, -$z(Cl)$).
Here we stick to the notation for the experimental structure.
 }

\label{ZrNCl-struct}
\begin{tabular}{|l|c|c|c|c|c|c|c|c|c|c|}
\hline
Compound          & $a$      & $c$      & $z(Zr/Hf)$ & $z(N)$  & $z(Cl)$ & exp.  \\
\hline
ZrNCl             & 3.6046   & 9.224    & 0.3577    & 0.5931 & 0.1634  & Ref.~\onlinecite{HfNCl}  \\
Li$_{0.25}$ZrNCl  & 3.591    & 9.280    & 0.6379    & 0.4086  & 0.1655  & Ref.~\onlinecite{LiZrNCl} \\
HfNCl             & 3.5767   & 9.237    & 0.3585     & 0.5928  & 0.1639 & Ref.~\onlinecite{HfNCl}  \\
Na$_{0.25}$HfNCl  & 3.5879   & 9.8928   & 0.6309     & 0.4110  & 0.1676  & Ref.~\onlinecite{NaHfNCl} \\
\hline
\end{tabular}
\end{table}

\begin{figure}[htb]
\includegraphics[width=0.95\linewidth]{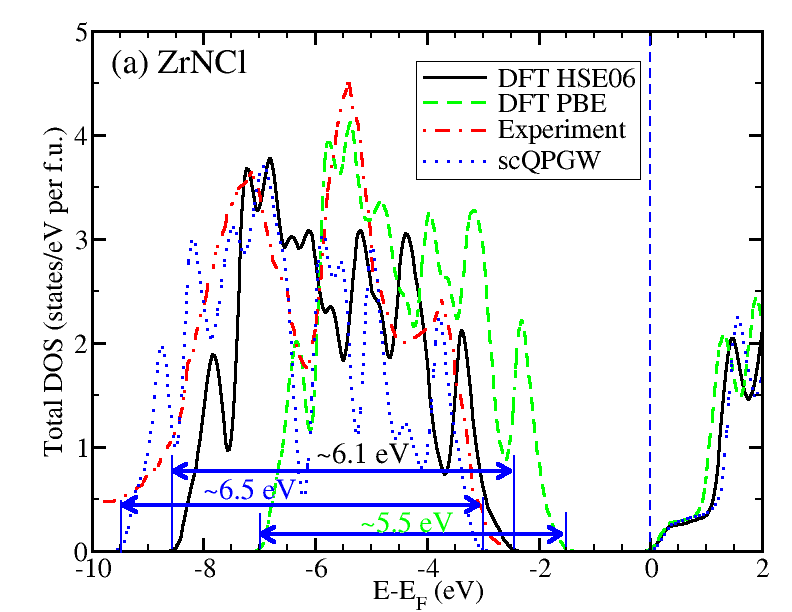}
\includegraphics[width=0.95\linewidth]{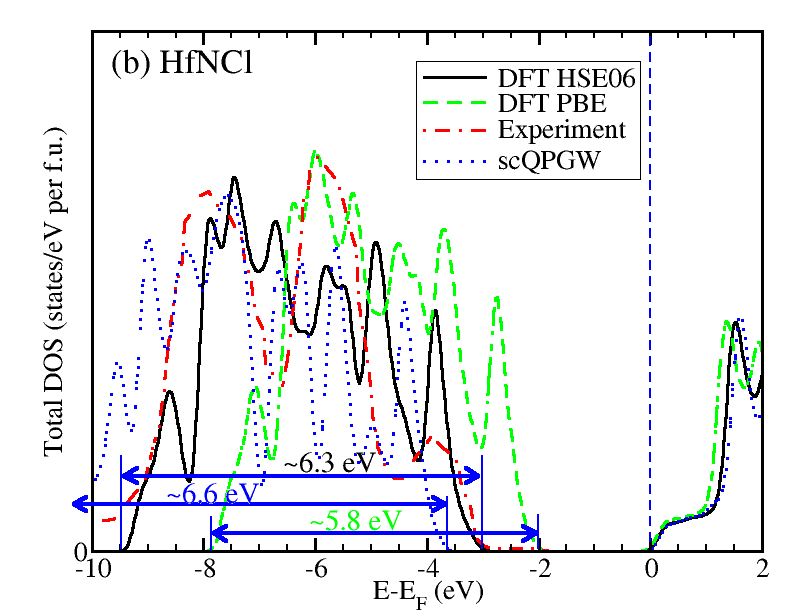}
\caption{
\textbf{Density of states of (a)ZrNCl and (b)HfNCl.}
The theoretical total DOS is calculated using the scQPGW method and
the PBE functional and HSE06 hybrid functional within DFT .
The experimental data is the photoemission spectra taken from Ref.~\onlinecite{PES-HfNCl}. }
\label{DOS-ZrNCl}
\end{figure}

\begin{figure}[htb]
\includegraphics[width=0.95\linewidth]{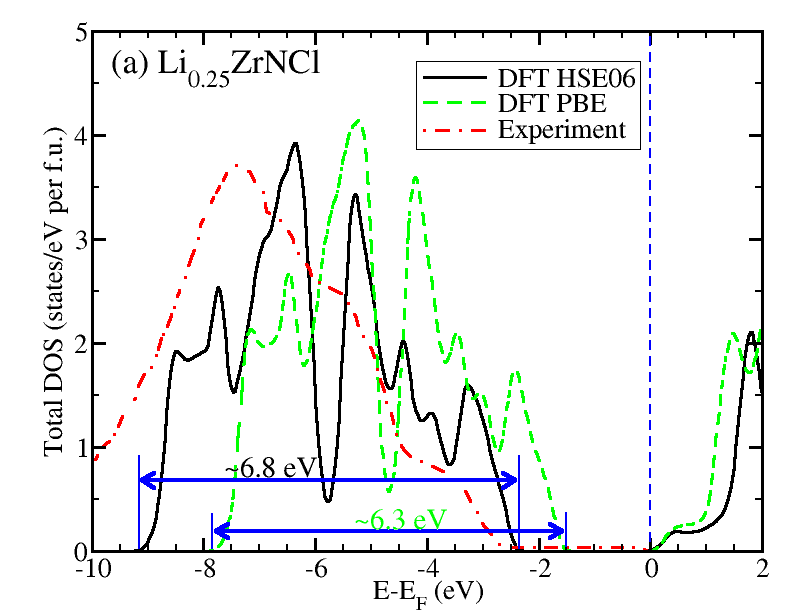}
\includegraphics[width=0.95\linewidth]{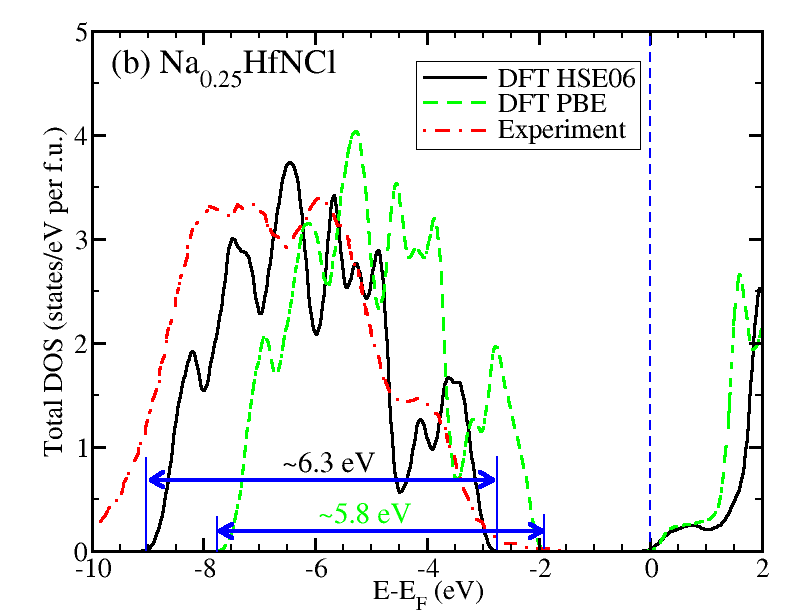}
\caption{
\textbf{Density of states of (a)Li$_{0.25}$ZrNCl and (b)Na$_{0.25}$HfNCl.}
The theoretical total DOS is calculated using
both the PBE functional and HSE06 hybrid functional. Doping is approximated by placing two
Li/Na atoms at (0, 0, 0) and (0.5, 0.5, 0) in the 2$\times$2$\times$1 supercell.
The experimental data is the photoemission spectra taken from Ref.~\onlinecite{PES-LiZrNCl}
for Na$_{0.42}$ZrNCl and Ref.~\onlinecite{PES-NaHfNCl} for Na$_{0.22}$HfNCl.
 }
\label{DOS-LiZrNCl}
\end{figure}

\subsection{Density of States}

\begin{table*}[htb]
\caption{
The valence band widths and the fundamental band gaps in ZrNCl and HfNCl and their electron-doped compounds
calculated using the DFT-GGA, DFT-HSE06, and scQPGW methods and compared with experiments[\onlinecite{PES-ZrNCl, PES-HfNCl, PES-LiZrNCl, PES-NaHfNCl}].
The DFT-HSE06 reproduces accurately both the valence band widths and fundamental band gaps in both materials
whereas DFT-GGA underestimates the band width by 0.5-0.7 eV and the band gap by about 1.0 eV. 
scQPGW overestimates the valence bandwidth by 0.3-0.4 eV and the band gap by 0.5 eV.
 }

\label{bandwidth-and-gap}
\begin{tabular}{|c|l|c|c|c|l|}
\hline
           & Compounds    & DFT-GGA & DFT-HSE06 & scQPGW  & exp. \\
\hline
Valence band width & ZrNCl        & 5.5     & 6.1       & 6.5 &6.1 (Ref.~\onlinecite{PES-HfNCl}) \\
           & Li$_{0.25}$ZrNCl & 6.3 & 6.8       &  &7.0 (Ref.~\onlinecite{PES-LiZrNCl} for Na$_{0.42}$ZrNCl) \\
           & HfNCl        & 5.8     & 6.3       & 6.6 &6.3 (Ref.~\onlinecite{PES-HfNCl})  \\
           & Na$_{0.25}$HfNCl & 5.8 & 6.3       &  &6.3 (Ref.~\onlinecite{PES-NaHfNCl} for Na$_{0.22}$HfNCl) \\
\hline
Fundamental band gap   & ZrNCl        & 1.6     & 2.5       & 3.1 &2.5 (Ref.~\onlinecite{PES-ZrNCl}) \\
           & HfNCl        & 2.0     & 3.1       & 3.7 &3.2-3.4 (Ref.~\onlinecite{PES-HfNCl}) \\
\hline
\end{tabular}
\end{table*}

The scQPGW, DFT-HSE06 and DFT-GGA calculated total density of states for the parent ZrNCl and HfNCl compounds are shown in
Fig.~\ref{DOS-ZrNCl} and compared to available experiments reported in Ref.~\onlinecite{PES-HfNCl}.
While the overall shape of the DOS is similar in all the methods, the bandwidths of the valence states and the band gap
between the valence states and conduction states differ substantially. We collect the valence bandwidths and the fundamental band gaps in
Table~\ref{bandwidth-and-gap} and compare to experimental values reported in Refs.~\onlinecite{PES-HfNCl, PES-ZrNCl}.
Clearly HSE06 account very well the valence bandwidths and band gaps in both materials, whereas GGA (scQPGW)
underestimates (overestimates) the band widths by 0.5-0.7 eV (0.3-0.4 eV) and the fundamental band gaps by about 1.0 eV (0.5 eV).

Upon electron doping, except a shift of the chemical potential (Fermi energy), the overall DOS of Na$_{0.25}$HfNCl remains
unchanged, as shown in Fig.~\ref{DOS-ZrNCl}(b), in consistent with experimental observations reported in Ref.~\onlinecite{PES-NaHfNCl}.
Here HSE06 functional again reproduces correctly the valence band width and the gap between the valence bands and conduction bands
whereas GGA underestimates both quantities to similar magnitudes as in the parent compound. On the other hand, the calculated
valence band widths in Li$_{0.25}$ZrNCl increases by about 0.7-0.8 eV compared to the parent compound ZrNCl, as shown in Fig.~\ref{DOS-ZrNCl}(a)
and Table~\ref{bandwidth-and-gap}.  This observation is consistent with the trend seen in experiment~[\onlinecite{PES-LiZrNCl}]
where the valence band widths in Na$_{0.42}$ZrNCl increases to about 7.0 eV. Note that HSE06 again gives better estimation of the valence band width than
GGA in Li$_{0.25}$ZrNCl.

As a result, the electronic structures in both the parent and the electron doped ZrNCl and HfNCl are much better accounted for by DFT-HSE06 than DFT-GGA.

\begin{table*}[htb]
\caption{
Phonon frequencies of the $E_g$ mode at zone center $\Gamma$ point
and zone boundary $K$ point in the parent ZrNCl and HfNCl compounds
and electron-doped Li$_{0.25}$ZrNCl and Na$_{0.25}$HfNCl compounds
calculated by frozen phonon calculations in the DFT-GGA and DFT-HSE06 framework.
For comparison, the corresponding phonon frequencies from LRT-LDA calculation in Ref.~\onlinecite{ZrNCl},
SCDFT-RPA calculations in Ref.~\onlinecite{SCDFT}, as well as from two experimental measurements
(exp.1 from Ref.~\onlinecite{phonon1} and exp.2 from Ref.~\onlinecite{phonon2}) are also displayed.
For electron doping, two Li/Na atoms are placed at (0, 0, 0) and (0.5, 0.5, 0) in the
2$\times$2$\times$1 supercell of the original unit cell.
Notice the strong phonon softening of the N-N $E_g$ mode near K point upon electron doping.
This phonon softening is strongly enhanced in DFT-HSE06 treatment.
 }

\label{Phonon-frequency2}
\begin{tabular}{|c|c|ccc|ccc|l|c|c|c|}
\hline
Compound           & mode   & \multicolumn{3}{|c|}{DFT-GGA}      & \multicolumn{3}{|c|}{DFT-HSE06}    & LRT-LDA   & SCDFT-RPA &  exp.1    & exp.2 \\
                   & $E_g$  & $\Gamma$  & $K$  & $\delta \omega$ & $\Gamma$  & $K$  & $\delta \omega$ & $\Gamma$  & $\Gamma$  & $\Gamma$  & $\Gamma$ \\
\hline
ZrNCl              & N-N    & 74.3      & 79.3 &  5.0            & 75.5      & 82.5 &  7.0            & 77.0      & 77.2      & 75.0      & 75.1 \\
Li$_{0.25}$ZrNCl   & N-N    & 77.1      & 69.6 & -7.5            & 76.7      & 61.9 & -14.8           & 74.4/78.6 & 74.8      & 75.5      &      \\
$\Delta\Omega$     & N-N    &  2.8      & -9.7 & -12.5           &  1.2      &-20.6 & -21.8           & -2.6/1.6  & -2.4      &  0.5      &      \\
\hline
HfNCl              & N-N    & 76.1      & 81.0 &  4.9            & 76.8      & 82.3 &  5.5            &           & 80.9      &           & 78.6 \\
Na$_{0.25}$HfNCl   & N-N    & 74.9      & 66.1 & -8.8            & 74.9      & 61.1 & -13.8           &           & 75.3      &           & 76.5 \\
$\Delta\Omega$     & N-N    & -1.2      &-14.9 &-13.7            & -1.9      &-21.2 & -19.3           &           & -5.6      &           & -2.1 \\
\hline
ZrNCl              & Zr-Zr  & 24.9      & 33.6 &  8.7            & 25.4      & 37.4 & 12.0            & 21.2      & 23.7      &  22.2     & 22.8 \\
Li$_{0.25}$ZrNCl   & Zr-Zr  & 25.8      & 31.1 &  5.3            & 25.3      & 27.1 &  1.8            & 21.0/20.5 & 22.5      &  22.1     &      \\
$\Delta\Omega$     & Zr-Zr  &  0.9      & -2.5 & -3.4            & -0.1      &-10.3 &-10.2            & -0.2/-0.7 & -1.2      &  -0.1     &      \\
\hline
HfNCl              & Hf-Hf  & 18.2      & 24.4 &  6.2            & 18.7      & 25.1 &  6.4            &           & 20.6      &           & 19.4 \\
Na$_{0.25}$HfNCl   & Hf-Hf  & 18.2      & 21.4 &  3.2            & 18.5      & 21.4 &  2.9            &           & 17.1      &           & 19.5 \\
$\Delta\Omega$     & Hf-Hf  &  0.0      & -3.0 & -3.0            & -0.2      & -3.7 & -3.5            &           & -3.5      &           &  0.1 \\
\hline
\end{tabular}
\end{table*}

\begin{table}[htb]
\caption{
Phonon frequencies of the $E_g$ mode at zone center $\Gamma$ point
and zone boundary $K$ point in ZrNCl and HfNCl compounds with structures
with/without electron doping
calculated by frozen phonon calculations in the DFT-GGA and DFT-HSE06 framework.
Notice that upon electron doping, the changes in the crystal structure don't
cause substantial changes in the phonon frequencies.
Therefore, the phonon softening shown in Table~\ref{Phonon-frequency2} is
electronic in origin rather than structural.
 }

\label{Phonon-frequency3}
\begin{tabular}{|c|c|c|ccc|ccc|}
\hline
Compound           & structure used   & mode   & \multicolumn{3}{|c|}{DFT-GGA}      & \multicolumn{3}{|c|}{DFT-HSE06}  \\
                   &                  & $E_g$  & $\Gamma$  & $K$  & $\delta \omega$ & $\Gamma$  & $K$  & $\delta \omega$ \\
\hline
ZrNCl              & ZrNCl            & N-N    & 74.3      & 79.3 &  5.0            & 75.5      & 82.5 &  7.0  \\
                   & Li$_{0.25}$ZrNCl & N-N    & 76.7      & 81.4 &  4.7            & 77.5      & 83.0 &  5.5  \\
$\Delta\Omega$     &                  & N-N    &  2.4      &  2.1 & -0.3            &  2.0      &  0.5 & -1.5  \\
\hline
HfNCl              & HfNCl            & N-N    & 76.1      & 81.0 &  4.9            & 76.8      & 82.3 &  5.5  \\
                   & Na$_{0.25}$HfNCl & N-N    & 75.4      & 80.5 &  5.1            & 75.9      & 81.6 &  5.7  \\
$\Delta\Omega$     &                  & N-N    & -0.7      & -0.5 &  0.2            & -0.9      & -0.7 &  0.2 \\
\hline
ZrNCl              & ZrNCl            & Zr-Zr  & 24.9      & 33.6 &  8.7            & 25.4      & 37.4 & 12.0  \\
                   & Li$_{0.25}$ZrNCl & Zr-Zr  & 26.1      & 34.6 &  8.5            & 26.5      & 35.4 &  8.9  \\
$\Delta\Omega$     &                  & Zr-Zr  &  1.2      &  1.0 & -0.2            &  1.1      & -2.0 & -3.1  \\
\hline
HfNCl              & HfNCl            & Hf-Hf  & 18.2      & 24.4 &  6.2            & 18.7      & 25.1 &  6.4  \\
                   & Na$_{0.25}$HfNCl & Hf-Hf  & 18.1      & 24.3 &  6.2            & 18.5      & 25.0 &  6.5  \\
$\Delta\Omega$     &                  & Hf-Hf  & -0.1      & -0.1 &  0.0            & -0.2      & -0.1 &  0.1  \\
\hline
\end{tabular}
\end{table}

\begin{figure}[htb]
\includegraphics[width=0.99\linewidth]{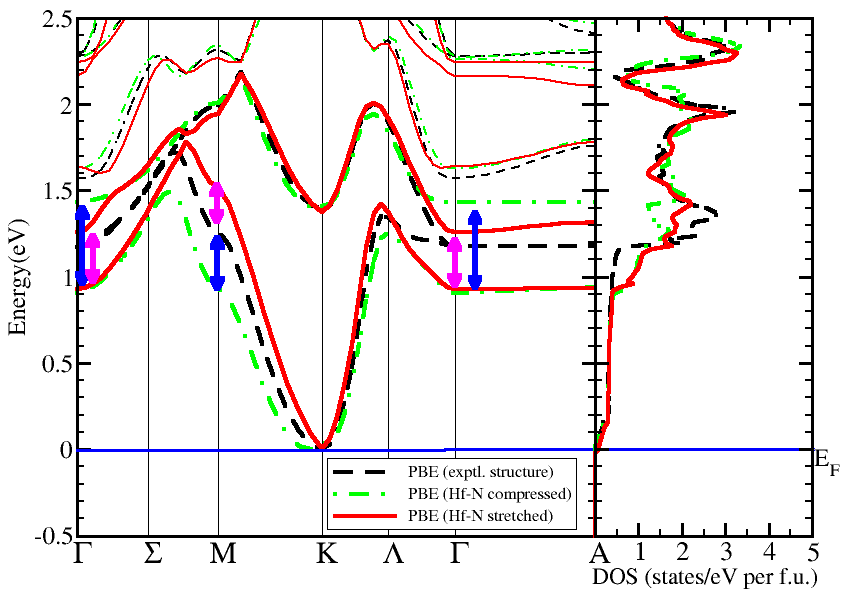}
\includegraphics[width=0.99\linewidth]{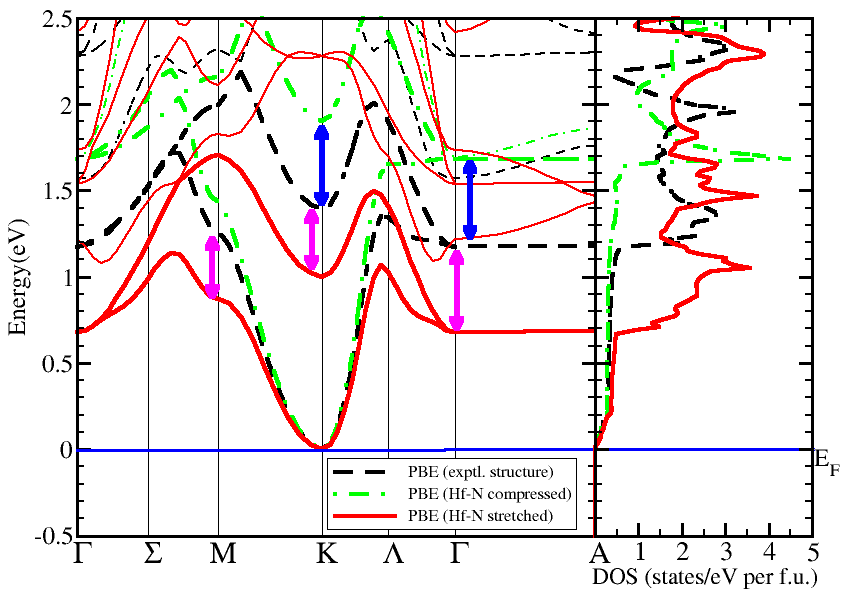}
\caption{
\textbf{Band structure and density of states of $\beta$-HfNCl.}
The band structure and density of states of $\beta$-HfNCl calculated
using the PBE functional with the experimental crystal structure[\onlinecite{HfNCl}]
and with the experimental crystal structure except the Hf and/or N atoms are shifted
in the $ab$ plane (top panel) and along the $c$-axis (bottom panel)
such that the Hf-N bond is compressed or stretched by 0.09 $\AA$.
For simplicity, the different layer stacking along $c$-axis is ignored.
Note the Fermi level is set to the bottom of the conduction band since we
are interested in electron doping.
}
\label{MNX-band}
\end{figure}

\subsection{Phonon frequencies}

It was suggested\cite{ZrNCl} that the in-plane optical phonons ($E_g$ mode) from
vibrations of N and Zr atoms dominate the EPC in Li$_{1/6}$ZrNCl.
We therefore compute the phonon frequencies at $\Gamma$ and $K$ points of two $E_g$ modes involving the in-plane vibration
of N and Zr/Hf atoms as shown in Fig.\ref{struct-ZrNCl} in ZrNCl, Li$_{0.25}$ZrNCl, HfNCl and Na$_{0.25}$HfNCl
using the frozen phonon method with both GGA and HSE06 functionals.
The structural parameters used in our calculations are the same as in Table \ref{ZrNCl-struct}.
We using a 2$\times$2$\times$1 supercell to calculate the phonon frequencies at $K$ points.
The phonon frequencies in Li$_{0.25}$ZrNCl (Na$_{0.25}$HfNCl) are calculated using a 2$\times$2$\times$1 supercell where
two Li (Na) atoms are placed at (0, 0, 0) and (0.5, 0.5, 0) in the supercell.
The calculated phonon frequencies are shown in Table \ref{Phonon-frequency2}
and compared with previous LRT-LDA calculation [\onlinecite{ZrNCl}] and SCDFT-RPA calculation[\onlinecite{SCDFT}]
and available experiments~[\onlinecite{phonon1, phonon2}].
A few points can be drawn from our calculations:\\
1)The phonon frequencies at $\Gamma$ points are similar in HSE06 and GGA and close to experimental values,
in consistent with other theoretical studies[\onlinecite{ZrNCl, SCDFT}].\\
2)Upon electron doping, both $E_g$ phonon modes strongly softens along the $\Gamma$-$K$ direction.
For the $E_g$ phonon mode involving in-plane nitrogen vibration,
the HSE06 (GGA) calculated phonon frequency at $K$ point softens by 29/25 (16/18) percent
or 21.8/19.3 meV (12.5/13.7 meV) compared to $\Gamma$ point
in ZrNCl/HfNCl upon 0.25/f.u. electron doping.
The corresponding numbers are 40/19 (13/17) percent or 10.2/3.5 meV (3.4/3.0 meV) for the $E_g$ mode involving the in-plane Zr/Hf vibration.\\
3)The observed softening is not due to a change of the crystal structure induced by doping.
As shown in Table~\ref{Phonon-frequency3},
we calculate the phonon frequencies with the crystal structures of Li$_{0.25}$ZrNCl and Na$_{0.25}$HfNCl but without doping the Li/Na atoms,
and find they are similar to the corresponding values in the parent compounds of ZrNCl and HfNCl, respectively,
i.e., we don't observe phonon softening at $K$ point. Hence, the phonon softening is of electronic rather than structural origin.

Therefore we conclude that for the $E_g$ phonon modes,
1) strong EPC occurs around the zone boundary $K$ point but not the zone center $\Gamma$ point,
consistent with the LRT-LDA results reported in Ref.~\onlinecite{ZrNCl},
and 2) HSE06 has much stronger EPC around $K$ point than GGA. Our results suggest that while
LDA/GGA produces reasonable phonon frequencies near $\Gamma$ point, it overestimates
the phonon frequencies around the zone boundary.  This is consistent with the comparison of the LRT-LDA phonon DOS
with the experimental one in Ref. \onlinecite{ZrNCl} and can be tested by future experiments.

\subsection{Band structure and electron-phonon coupling}

The DFT-PBE band structure and density of states of $\beta$-HfNCl is shown in Fig.\ref{MNX-band},
which is calculated with its experimental crystal structure[\onlinecite{HfNCl}]
without/with Hf and/or N atoms shifted in the $ab$ plane and along the $c$-axis to change the Hf-N bonding length.
In the calculations, the experimental stacking of the layers along the $c$-axis is ignored for simplicity.
Note that the study of Ref.\onlinecite{ZrNCl} used quite small electron
doping (1/6), our calculations suggest at somewhat higher
electron-doping the optical vibration of N and Zr atoms along $c$
direction are also relevant to EPC.
As the arrows indicated in the left panel of Fig.\ref{MNX-band}, there are large REPMEs
for the first two conduction bands,
which are very flat along the $\Gamma$-A symmetry line,
resulting in high density of states as shown in the right panel of Fig.\ref{MNX-band}.
Such high density of states makes electron-doping hardly change
the chemical potential in this region.

The study of Ref.\onlinecite{ZrNCl} suggests the EPC are mainly
contributed by in-plane optical phonons ($E_g$ mode) from
vibrations of N and Zr atoms, with strong EPC near the zone
boundary $K$ point. We have shown that the experimental electronic structures
of both parent and electron doped ZrNCl and HfNCl are much better
reproduced by HSE06 than LDA. We also find that the phonon
frequencies of the $E_g$ mode soften more strongly in HSE06 than
LDA near $K$ point, similar to the oxygen breathing mode near the
zone boundary points in (Ba,K)BiO$_3$. The stronger softening of
the HSE06 $E_g$ phonon frequencies is consistent with the
comparison of the experimental and LRT-LDA phonon DOS shown in
Ref.~\onlinecite{ZrNCl}. It is therefore appropriate to use our
approach (equation \ref{lratio1}) to estimate the realistic EPC
in this family. In the electron-doped $\beta$-ZrNCl compound, the
REPMEs of the conduction bands 
associated with Zr-N
vibrations in the plane and along $c$ direction are substantially
enhanced from 2.9-4.0 eV/$\AA$ in LDA to 3.9-4.7 eV/$\AA$ in
HSE06. The enhanced REPME in HSE06 results in $\lambda_H$ $\sim$
0.8 and T$_c$=18 K (see Table~\ref{deformation-and-Tc}) , in
agreement with experimental value of 16.0 K.
Interestingly, the enhanced electron-phonon coupling evaluated by DFT-HSE06 
doesn't lead to a larger specific heat coefficient $\gamma=\frac{2}{3}(1+\lambda)\pi^2k_B^2N(0)$ compared 
to its DFT-LDA value because the electronic density of states at Fermi level, 
i.e. N(0), is reduced 
in DFT-HSE06 by about 20$\%$ compared to DFT-LDA at the same time.

\subsection{Material and doping dependence of T$_c$}

Our theory explains several counterintuitive experimental
observations. It explains naturally the higher T$_c$ in electron-doped
$\beta$-HfNCl than electron-doped $\beta$-ZrNCl, despite the fact that the
Hf atom is substantially heavier than Zr atom and therefore it
would be expected to have lower frequency and lower T$_c$ in a BCS
picture. It turns out that the dominant factor governing the T$_c$
is the value of the REPME which is substantially
larger in $\beta$-HfNCl than in $\beta$-ZrNCl, and
increased from 3.8-4.4 eV/$\AA$ in LDA to 5.1-5.3 eV/$\AA$ in
HSE06. Since $\beta$-HfNCl is isostructural to $\beta$-ZrNCl, the
enhanced REPME give $\lambda_H$ $\sim$ 1.1 and
T$_c$=25 K for electron-doped $\beta$-HfNCl (see Table~\ref{deformation-and-Tc}),
in good agreement with experiment.
The 16.5 K superconductivity in electron-doped
$\alpha$-TiNCl[\onlinecite{TiNCl}] may also be well accounted for
in the same way.

Another mystery in this family is the weak doping dependence of
T$_c$, which is quite different from the trends in other known
high temperature superconductors, including the BaBiO$_3$-based ones
where T$_c$ depends crucially on the doping level.
To understand this, notice that in this family the
conduction band as shown in Fig.\ref{MNX-band} has the largest REPMEs thus strong EPC
at about 1.0 eV above the bottom of the conduction band.
In the same energy region,
this conduction band, being flat in certain parts of the Brillouin zone (along $\Gamma-A$ for example),
results in high density of states
which makes the Fermi level hardly change with increasing doping level after
some critical doping.
The combination of the large REPMEs and high density of states in the same energy region is likely
a plausible explanation of the weak doping dependence of T$_c$ in the HfNCl family.

\section{Summary and discussion}

\subsection{Summary}
\begin{table*}[htb]
\caption{
REPME $D$ (eV/$\AA$) for the most important vibration mode, the
total EPC $\lambda$, average phonon frequency $\omega_{log}$
(K) and calculated T$_c$ (K) in LDA and HSE06 are
displayed with experimental T$_c$ for selected compounds at
optimal electron or hole doping.
$\omega_{log}$ in HSE06 is estimated from the corresponding
LRT-LDA value as described in Section III.
Critical temperatures in LDA and HSE06 are obtained from
eq.~(\ref{eq3}) using $\mu^*$=0.10. (We concentrate on the dominating 
effect of $\lambda$ and ignore the small variations in the values of $\mu^*$ 
in the considered materials,  
which in practice affects the calculated T$_c$ by a few Kelvins but doesn't affect 
our conclusions.)
We approximate the LDA $\omega_{log}$ for
HfNCl to be about 80$\%$ of the value for ZrNCl according to the
mass of Hf and Zr and their contribution to the total $\lambda$
(about 40$\%$ according to Ref.\onlinecite{ZrNCl}).
 }

\label{deformation-and-Tc}
\begin{tabular}{|c|c|c|c|c|c|c|c|c|c|c|}
\hline
Compounds    & mode   & $D$ (LDA)   & $D$ (HSE)   & $\lambda$(LDA) &  $\lambda$ (HSE) & $\omega_{log}$ (LDA) & $\omega_{log}$(HSE) &T$_c$(LDA)  & T$_c$ (HSE) & T$_c$ (exp.) \\
BaBiO$_3$    & O breathing at $R$ & 7.6         & 13.3        & 0.33           &  1.0            & 550 & 450          & 0.6   & 31        & 32.0 [\onlinecite{Cava}] \\
BaPbO$_3$    & O breathing at $R$ & 10.1        & 11.2        & 0.58           &  0.72            & 500 & 480          & 10.3   & 18        & -- \\
Ba$_3$Bi$_2$O$_7$  & O breathing at $R$ & $\sim$0  & $\sim$0  & $\sim$0        &  $\sim$0         &  -- & --    & $\sim$0  & $\sim$0     & $<$2 [\onlinecite{BBO327}] \\
ZrNCl        & $E_g$ at $\Gamma$ (in-plane)   & 2.9-4.0         & 3.9-4.7         & 0.52           &  0.8            & 422 & 390         & 6.0   & 18        & 16 [\onlinecite{MNX}] \\
HfNCl        & $E_g$ at $\Gamma$ (in-plane)  & 3.8-4.4         & 5.1-5.3         &                &  1.1            & 340 & 310         & --   & 25        & 25.5 [\onlinecite{Yamanaka}] \\
\hline
\end{tabular}
\end{table*}

Table \ref{deformation-and-Tc} summarizes our results on selected lattice dynamical properties (REPMEs, $\omega_{log}$), 
the EPC and the T$_c$ within the Migdal-Eliashberg theory for the materials studied above. 
Enhancements of the REPMES in the more advanced and accurate HSE06 screened hybrid functional treatment and scQPGW method  
compared to LDA give rise to stronger EPC strengths and softening of strongly electron-phonon coupled phonon modes, 
consistent with many experimental observations as discussed above.  
The enhanced EPC and renormalized phonon frequency $\omega_{log}$ readily account for 
the rather high temperature superconductivity in the doped bismuthates and transition metal chloronitrides within the Migdal-Eliashberg theory. 
In addition, using the important REPMEs, we can explain the material dependence of superconductivity, such as high T$_c$ (up to 32 K) in (Ba,K)BiO$_3$, 
intermediate T$_c$ ($<$20 K) in (Ba,La)PbO$_3$ and Ba(Pb,Bi)O$_3$ and low T$_c$ ($<$2K) in (Ba,K)$_3$(Bi,Pb)$_2$O$_7$. 
We are also able to explain the material and doping dependence of T$_c$ in $A_x$HfNCl and $A_x$ZrNCl ($A$=Li, Na, etc.)

\subsection{Two general sources of the underestimation of EPI by LDA}

The underestimation of the EPI by LDA in materials close to a metal-insulator transition is very general, 
and is closely related to the failure of LDA in describing the ground state properties of the parent compounds.
This can be understood intuitively as follows: LDA overestimates dielectric screening while the EPI is inversely proportional to
the dielectric constant\cite{Allen3}, therefore the EPI strength is reduced in
LDA relative to its true value.

The materials studied in this paper represent two general sources which cause the underestimation of the EPI by LDA.
The first general source is the underestimation of the relevant band widths by LDA which causes an underestimation of the REPMEs.
The transition metal chloronitrides represent this case where LDA underestimates the band width of the lowest conduction band by about 10$\%$ in the 
parent compounds and 20$\%$ in the 1/4 electron doped compounds compared the HSE06 treatment. 
(see Fig.\ref{DOS-ZrNCl} and \ref{DOS-LiZrNCl} for the first peak positions in the DOS of the conduction bands from GGA, HSE06, and/or scQPGW.)
The second general source is the underestimation of the fundamental band gap by LDA, where the fundamental band gap is caused by 
atomic distortions. In this case the fundamental band gap is closely connected to the REPMEs of the corresponding atomic distortions.
The bismuthates belong to the second case where the fundamental band gap is of indirect type and is only about 0.15 eV in LDA\cite{Meregalli} 
while the experimental value is in the range of 0.5-0.9 eV\cite{BBO-gap}. Notice that the corresponding value in the HSE06 treatment is about 0.65 eV.\cite{hybrid1}

\subsection{Screened and unscreened hybrid functionals: HSE06 vs B3LYP}
There is much doubt on the applicability of hybrid functional to metallic materials 
partially because the famous B3LYP type hybrid functional fails in 
describing metallic systems. One reason is that the B3LYP functional fails to attain the exact homogeneous electron gas
limit\cite{Hybrid-review1} due to the spin density wave instability in the Hartree-Fock method\cite{Overhauser}. 
Paier, Marsman, and Kresse found that observing this limit is exceedingly important for extended
systems.\cite{HSE-review, Hybrid-review1} On the other hand, the HSE-type \textit{screened} functional has been designed to 
describe the homogeneous electron gas exactly therefore the HSE-type functional performs much better than the B3LYP 
functional.\cite{Hybrid-review1} Detailed comparisons of the performance of B3LYP and HSE-type functionals and other functionals 
are available in the literature (see for example, Refs.~\onlinecite{Hybrid-review1} and \onlinecite{Hybrid-review2}). 
For lattice dynamical properties, it was found that the B3LYP functional 
overestimate the EPI of some phonon modes in extended systems such as graphene and graphite,\cite{graphene} 
whereas in the molecular systems such as C$_{60}$, both the B3LYP and HSE-type functionals show similar 
improved estimation of the EPI.\cite{C60, C602} 
Therefore, in the present paper we choose the \textit{screened} hybrid functional HSE06 to study 
the doped bismuthates and transition metal chloronitrides, as they are extended metallic materials.
The HSE06 functional is proven to be a good choice as it reproduces accurately the electronic structures and phonon frequencies 
of the considered materials.

\subsection{Screening parameter}
In the construction of the HSE-type screened hybrid functional, it includes 1/4 of screened non-local exchange potential whose range 
is determined by a range-separation parameter $\mu$, where $\mu$=0, 0.2, and $\infty$ corresponds to PBE0, HSE06, and PBE functional, respectively.\cite{Hybrid-review2} 
In more metallic materials, the HSE06 functional tends to overestimate the band widths and might perform worse than the conventional LDA functional.\cite{Hybrid-review2}
To correct this issue of the HSE06 functional, the range-separation parameter $\mu$ in the HSE-type functional has to be adjusted 
so as to get better agreements with the experimental or QPGW electronic structures in the normal state. 
We note that in the strongly overdoped region of the bismuthates and transition metal chloronitrides, 
the overscreening problem of LDA becomes less severe and
the screening parameter in HSE06 functional has to be adjusted.
Since the QPGW method can determine the degree of non-locality of the exchange-correlation potential self-consistently, 
the $\mu$ parameter in the screened hybrid functional can be tuned to fit the QPGW results. 
In this way, the $\mu$ parameter in the screened hybrid functional can be determined by the QPGW method and is no longer an empirical parameter.
Here we take MgB$_2$ as an example to demonstrate this.
We show below that corrections beyond LDA/GGA are needed to better describe experimental observations in MgB$_2$.
By adjusting $\mu$ in the HSE-type functional to fit the QPGW results, a modified HSE functional
fixes the discrepancies between LDA/GGA and experiments and results in slightly stronger EPC strength and
larger phonon linewidths, in consistent with the fact that LDA/GGA functional in average underestimates the experimental
phonon line widths~[\onlinecite{MgB2-linewidth1, MgB2-linewidth2, MgB2-linewidth3, MgB2-linewidth4}].
Therefore, the modified HSE functional provides a better estimation of the EPC strength in MgB$_2$.

\begin{figure}[htb]
\includegraphics[width=0.99\linewidth]{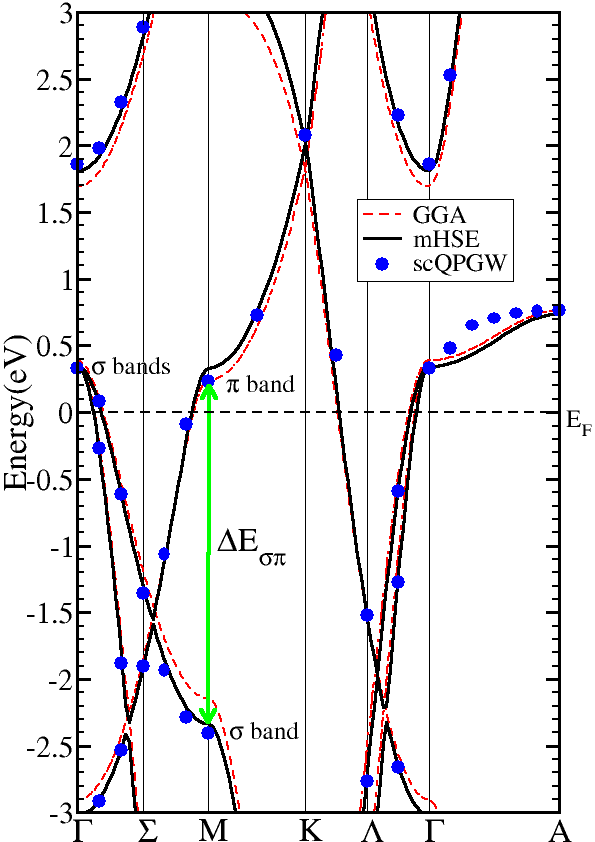}
\caption{
\textbf{Band structure of MgB$_2$ calculated by DFT-GGA and DFT-mHSE06, and QPGW }
The DFT-mHSE ($\mu$=0.6) band structure agrees well with that calculated by scQPGW, 
with a splitting of the $\pi$ band and $\sigma$ band at $M$ point (i.e.,$\Delta E_{\sigma\pi}$) about 2.6 eV, 
in good agreement with experiments\cite{MgB2-optics}. 
DFT-GGA underestimates this splitting $\Delta E_{\sigma\pi}$ by about 0.2 eV and overestimate the size of the $\sigma$ Fermi surface centered at $\Gamma$ point. 
}
\label{MgB2-band}
\end{figure}

\begin{table*}[htb]
\caption{Various quantities in MgB$_2$ calculated by DFT-GGA, DFT-HSE06, DFT-mHSE ($\mu=0.6$), scQPGW, and compared to available experiments, including
the first interband transition peak position (in eV) in optics, de Haas-van Alphen frequency (in T)
from the Fermi surface formed by the lower $\sigma$ band in the $\Gamma$ plane,
phonon frequency (in meV) of the $E_{2g}$ mode,
calculated reduced electron-phonon matrix electron (in eV/$\AA$) of the $E_{2g}$ mode.
 }

\label{MgB2}
\begin{tabular}{|l|c|c|c|c|l|l|}
\hline
                                                     & DFT-GGA   & DFT-HSE06 & DFT-mHSE  & scQPGW & exp.                                     & other DFT-LDA/GGA \\
                                                     & $\mu$=$\infty$ & $\mu$=0.2 & $\mu$=0.6 &     &                                     &  \\
\hline
peak position $\Delta E_{\sigma\pi}$ (eV)            & 2.4       & 3.0       & {\bf 2.6}   & 2.6 & {\bf 2.6} (Ref.~\onlinecite{MgB2-optics})     & 2.4 (Ref.~\onlinecite{MgB2-optics-theo}) \\
de Haas-van Alphen frequency $F_{\sigma\Gamma}$ (T)  & 730       & 350       & {\bf 580 }  &  & {\bf 535-546} (Ref.~\onlinecite{MgB2-dehaas}) & 728-878 (Ref.~\onlinecite{dHvA-theo1, dHvA-theo2, dHvA-theo3}) \\
phonon frequency $\omega_{E_{2g}}$ (meV)             & 71.8      & 66.0      & {\bf 70.5}  &  & {\bf 69.6-71.2} (Ref.~\onlinecite{MgB2-raman}) & 58-83 (Refs.~\onlinecite{MgB2-E2g1, MgB2-E2g2, MgB2-E2g3}) \\
REPME of $E_{2g}$ mode (eV/$\AA$)                    & 12.3      & 16.5      & 13.9        &  &                                        & 13 (Ref.~\onlinecite{MgB2-pickett}) \\
\hline
\end{tabular}
\end{table*}

While LDA/GGA is commonly considered to be accurate for MgB$_2$, there are various experimental observations are not well reproduced by LDA/GGA.
For example, it was shown in Ref.~\onlinecite{MgB2-dehaas} that LDA/GGA overestimated several de Haas-van Alphen frequencies by 200 T.
For the smaller Fermi surface formed by the $\sigma$ band in the $\Gamma$ plane, LDA/GGA overestimated its size by 30 percent.[\onlinecite{MgB2-dehaas}]
In the optical conductivity, the peak of the interband transition experimentally at about 2.6 eV was estimated to locate at about 2.4 eV by DFT-LDA/GGA.[\onlinecite{MgB2-optics}]
That is to say, LDA/GGA underestimates the splitting between the lower $\sigma$ band and the upper $\pi$ band.
These overestimations and underestimations of experimental observations by LDA/GGA suggest that corrections beyond LDA/GGA are needed to faithfully reproduce experimental results.

On the other hand, there are unexpectedly large discrepancies in the LDA/GGA results of the same quantity reported by different groups.
For example, the de Haas-van Alphen frequencies of the same Fermi surfaces calculated
by a few groups differ by 100-300 T.[\onlinecite{MgB2-dehaas}]
Another example, the phonon frequency of the most important $E_{2g}$ mode varies from 470-515 cm$^{-1}$[\onlinecite{MgB2-E2g1}] to
585 cm$^{-1}$[\onlinecite{MgB2-E2g2}] and to 575-665 cm$^{-1}$[\onlinecite{MgB2-E2g3}].
Note the differences in the values of the reported frequencies in the preprint and published
version of Ref.~\onlinecite{MgB2-E2g1} and Ref.~\onlinecite{MgB2-E2g3}.
These large discrepancies in the reported LDA/GGA results indicate unusual sensitive to the calculational details in MgB$_2$.

\begin{figure}[htb]
\includegraphics[width=0.99\linewidth]{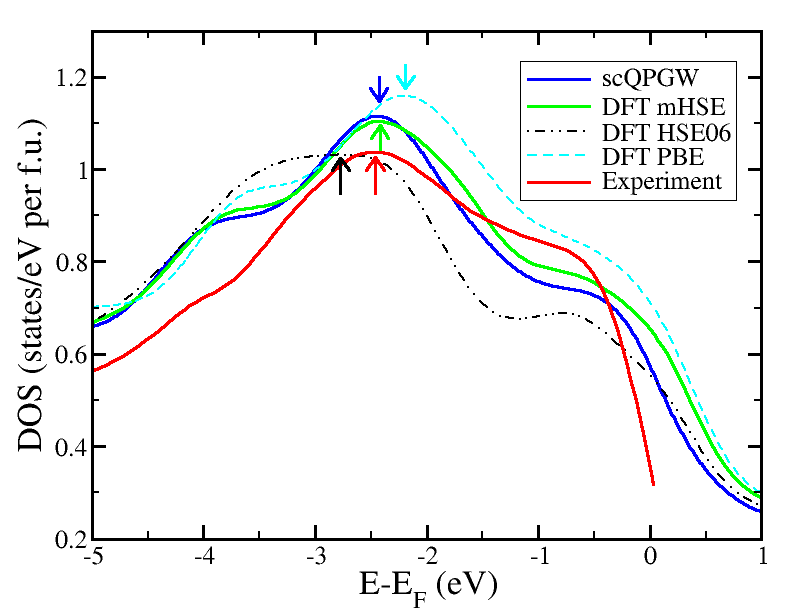}
\caption{
\textbf{Density of states of MgB$_2$ calculated by scQPGW, DFT-mHSE, DFT-HSE06, and DFT-GGA and compared to experiments\cite{PES-MgB2}. }
For each DOS of all the data, an arrow is drawn to indicate the peak position which is mainly contributed by the B $2p$ states.
The DOS of the scQPGW method, DFT-mHSE ($\mu$=0.6) and experiment\cite{PES-MgB2} 
share a common peak at about -2.4 eV whereas DFT-GGA underestimates this peak position by 0.2 eV (at about -2.2 eV) 
and DFT-HSE06 overestimates it by 0.4 eV (at about -2.8 eV). 
}
\label{DOS-MgB2}
\end{figure}

In the following, we correct the above discrepancies between LDA/GGA and experiments in the optical conductivity and de Haas-van Alphen frequencies by 
using HSE-type screened hybrid functional and QPGW method. The range of the non-local exchange in the HSE type functional 
is adjusted by tuning the $\mu$ parameter to fit the scQPGW results such as band structure (shown in Fig. \ref{MgB2-band}) and DOS (shown in Fig. \ref{DOS-MgB2}).
In the calculations, we use experimental structure (lattice constants $a=3.0834$ and $c=3.5213$, space group $P6/mmm$, \#191) from Ref.~\onlinecite{MgB2-struct}.

In Table~\ref{MgB2}, we collect a few quantities calculated by DFT-GGA, DFT-HSE06 and DFT-mHSE (a modified HSE functional with $\mu=0.6$), scQPGW, and compare them with
experimental values and previous theoretical calculations. The included quantities are the splitting $\Delta E_{\sigma\pi}$ between the lower $\sigma$ band
and the $\pi$ band around $M$ point (marked as an arrow in Fig.\ref{MgB2-band}) showing as a interband transition peak in the optical conductivity,
the de Haas-van Alphen frequency $F_{\sigma\Gamma}$ of the lower $\sigma$ band in the $\Gamma$ plane,
the phonon frequency $\omega_{E_{2g}}$ of the $E_{2g}$ mode, and the REPME (in eV/$\AA$) of the $E_{2g}$ mode.

Figure \ref{MgB2-band} shows the band structure of MgB$_2$ calculated by DFT-GGA, DFT-mHSE ($\mu=0.6$), and scQPGW.
Consistent with previous LDA/GGA calculations as shown in Table~\ref{MgB2}, our DFT-GGA calculation also underestimates the splitting $\Delta E_{\sigma\pi}$ by 0.2 eV
and overestimates the de Haas-van Alphen frequency $F_{\sigma\Gamma}$ by 190 T.
On the contrary, DFT-HSE06 overestimates $\Delta E_{\sigma\pi}$ by 0.4 eV and underestimates $F_{\sigma\Gamma}$ by 190 T, suggesting HSE06 overcompensates the
overscreening of LDA/GGA. Therefore we increase $\mu$ to reduce the range of the non-local exchange potential in HSE06 functional.
We find that a modified HSE type functional with $\mu$=0.6 simultaneously fit well to the scQPGW results and 
reproduces the experimental $\Delta E_{\sigma\pi}$, $F_{\sigma\Gamma}$ and $\omega_{E_{2g}}$
as shown in Table~\ref{MgB2}, Fig.~\ref{MgB2-band} and Fig.~\ref{DOS-MgB2}.  
As a result, MgB$_2$ is a material sits between conventional metals which are well described by LDA/GGA functional and conventional insulators which
are much better described by the HSE06 screened hybrid functional.

Since MgB$_2$ are better described by the mHSE functional with $\mu$=0.6, we now proceed to estimate the EPC strength in MgB$_2$ 
combining the LRT-LDA $\lambda$ and the frozen-phonon calculations. According to Ref.~\onlinecite{MgB2-E2g2}, 
the total LRT-LDA $\lambda$ is about 0.87, of which $\lambda_{\sigma}$=0.62 is contributed by the $E_{2g}$ mode of the $\sigma$ band
and $\lambda_{\pi}$=0.25 comes from the $\pi$ band.
Here we renormalize $\lambda_{\sigma}$ and ignore the difference in $\lambda_{\pi}$ 
because the latter makes a much smaller contribution to the total EPC. The estimated EPC strength in mHSE functional is
$\lambda_{mHSE}=(13.9/12.3)^2\lambda_{\sigma}+\lambda_{\pi}=0.79+0.25=1.04$, which is 0.17 bigger than the LRT-LDA value.
This is consistent with the fact that LDA/GGA functional in average underestimates the experimental
phonon line widths~[\onlinecite{MgB2-linewidth1, MgB2-linewidth2, MgB2-linewidth3, MgB2-linewidth4}]
and suggests that the mHSE functional provides a better estimation of the EPC strength in MgB$_2$.

In short, MgB$_2$ is closer to a conventional superconductor
than the bismuthates and transition metal chloronitrides. 
Applying the methodology we advocate in this paper, namely,
using a screened hybrid functional and/or GW method that describes well the normal state properties to obtain
an improved description of the superconducting state, does not lead to dramatic
modifications of the existing picture for MgB$_2$.

Although the difference between LDA/GGA and mHSE in MgB$_2$ is substantially smaller than in the case of (Ba,K)BiO$_3$,
it is consistent manifestation of the overscreening problem of the LDA/GGA functional.
It testifies our idea
that due to the overscreening problem, LDA/GGA underestimates the EPC in these compounds.
The realistic EPC thus has to be evaluated with a method which describes well the experimental
electronic structures and phonon frequencies. 
In addition to the computational more demanding GW method, 
the HSE06 \textit{screened} hybrid functional within DFT can serve as such a method for (Ba,K)BiO$_3$ and electron doped
ZrNCl and HfNCl while a modified HSE-type \textit{screened} hybrid functional within DFT is suitable for MgB$_2$.

\subsection{Other materials}
There is a class of materials where corrections beyond LDA/GGA are needed to 
account for the lattice dynamical properties and EPI.
In addition to the doped bismuthates and transition metal chloronitrides, 
it was noted that in graphene\cite{graphene} and organic molecular compounds,
such as $A_3$C$_{60}$\cite{C60, C602} and picene\cite{picene},
GW or (screened) hybrid DFT methods are needed to evaluate phonon-related quantities.
The latter two have narrow bands and may be in an antiadiabatic regime.\cite{C603}
This class may include other puzzling materials such as Tl$_x$PbTe\cite{fisher}
and undoubtedly many other systems still to be discovered.

\section{Conclusions}

In conclusion, we show in this paper that first principles
calculations based on LDA/GGA functional can significantly
underestimate the EPI strength in certain materials 
where there are large non-local correlations such as materials 
in the vicinity of a metal-insulator transition. The 
doped BaBiO$_3$ and HfNCl family are such materials-the parent compounds are insulators 
while the sufficiently doped compounds are metals. This underestimation of EPI is caused by 
the overscreening of LDA/GGA in insulators, semiconductors and low-carrier 
bad metals, which results in an underestimation of the relevant band widths 
and/or the fundamental band gaps-two general sources of the underestimation of electron phonon coupling. 

We present a simple but efficient method to evaluate 
the realistic EPC in these materials by combining 
the LDA linear response calculations and a few supercell calculations 
using a more advanced and accurate method such as GW and screened hybrid functional DFT 
which is able to provide a significantly improved description of normal state electronic 
structures and lattice dynamical properties of the materials compared to LDA.

We apply our method to evaluate the EPC in the doped 
BaBiO$_3$ and HfNCl family and find that the realistic electron phonon coupling 
are significantly enhanced over the LDA values and are strong enough to 
account for the rather high temperature superconductivity in these materials 
within the Migdal-Eliashberg theory. 
The puzzling discrepancies between theory and experiments in 
these materials are naturally resolved with our method. 
The ``other high-temperature superconductors" such as those summarized in Refs.\onlinecite{Pickett, pickett2} are
thus strongly coupled superconductors where the coupling of the
electrons to the lattice vibrations requires treatment of
correlations beyond LDA/GGA.

Our method is also important for rational design of new superconductors 
and other EPI related functional materials 
through first principles calculations because determining the electron phonon coupling 
reliably is an essential step. As an example, we predict up to 18 K 
superconductivity in the perovskite Ba$_{1-x}$La$_x$PbO$_3$ compound around $x=0.7$.

Our work suggests a critical reexamination of the realistic EPC
is necessary in these and other materials such as the correlated cuprates and iron-based superconductors.
Even in MgB$_2$ as discussed above, a careful reexamination shows LDA slightly underestimates the EPC.

Finally, we reiterate that our approach of estimating the realistic EPC can be improved by including 
more phonon modes and our work calls for implementing the linear response technique into 
GW and screened hybrid functional DFT to achieve a more accurate evaluation of the realistic EPC.  

\textit{Acknowledgments} We are grateful to Mac Beasley and Bob
Cava for renewing our interest in this problem, and to Sergey
Savrasov, Warren Pickett and Kristjan Haule for their useful
comments. This work was supported by the AFOSR-MURI program.

\end{document}